\documentclass[%
 reprint,
superscriptaddress,
 amsmath,amssymb,
 aps,
pra,
longbibliography,
floatfix,
]{revtex4-1}
\makeatletter
\renewcommand*\env@matrix[1][*\c@MaxMatrixCols c]{%
  \hskip -\arraycolsep
  \let\@ifnextchar\new@ifnextchar
  \array{#1}}
\makeatother

\usepackage{tensor}
\usepackage{graphicx}
\usepackage{dcolumn}
\usepackage{bm}
\usepackage{dsfont}
\usepackage{braket}
\usepackage{tikz}
\usetikzlibrary{matrix,decorations.pathreplacing,calc}
\usetikzlibrary{patterns}
\usepackage{xcolor}

\newcommand{\tikzmark}[1]{\tikz[overlay,remember picture,baseline=(#1.base)]
  \node (#1) {\strut};}
\newcommand{\tikznode}[2]{%
\ifmmode%
  \tikz[remember picture,baseline=(#1.base),inner sep=0pt] \node (#1) {$#2$};
\else
  \tikz[remember picture,baseline=(#1.base),inner sep=0pt] \node (#1) {#2};%
\fi}

\DeclareMathOperator{\spn}{span}
\DeclareMathOperator{\nll}{null}
\DeclareMathOperator{\rng}{range}

\begin{document}

\preprint{APS/123-QED}

\title{Analog spacetimes from nonrelativistic Goldstone modes in spinor condensates}

\author{Justin H.\ Wilson}
\affiliation{%
Institute of Quantum Information and Matter and Department of Physics,
Caltech, CA 91125, USA
}%
\affiliation{Department of Physics and Astronomy, Center for Materials Theory, Rutgers University, Piscataway, NJ 08854 USA}
\affiliation{Department of Physics and Astronomy, Louisiana State University, Baton Rouge, LA 70803, USA}
\affiliation{Center for Computation and Technology, Louisiana State University, Baton Rouge, LA 70803, USA}
\author{Jonathan B. Curtis}
\affiliation{Joint Quantum Institute and Condensed Matter Theory Center, Department of Physics, University of Maryland, College Park, Maryland 20742-4111, USA}
\author{Victor M.\ Galitski}
\affiliation{Joint Quantum Institute and Condensed Matter Theory Center, Department of Physics, University of Maryland, College Park, Maryland 20742-4111, USA}

\date{\today}

\begin{abstract}
It is well established that linear dispersive modes in a flowing quantum fluid behave as though they are coupled to an Einstein-Hilbert metric and exhibit a host of phenomena coming from quantum field theory in curved space, including Hawking radiation.
We extend this analogy to any nonrelativistic Goldstone mode in a flowing spinor Bose-Einstein condensate.
In addition to showing the linear dispersive result for all such modes, we show that the quadratically dispersive modes couple to a special nonrelativistic spacetime called a Newton-Cartan geometry. 
The kind of spacetime (Einstein-Hilbert or Newton-Cartan) is intimately linked to the mean-field phase of the condensate.
To illustrate the general result, we further provide the specific theory in the context of a pseudo-spin-1/2 condensate where we can tune between relativistic and nonrelativistic geometries.
We uncover the fate of Hawking radiation upon such a transition: it vanishes and remains absent in the Newton-Cartan geometry despite the fact that any fluid flow creates a horizon for certain wave numbers.
Finally, we use the coupling to different spacetimes to compute and relate various energy and momentum currents in these analog systems.
While this result is general, present day experiments can realize these different spacetimes including the magnon modes for spin-1 condensates such as $^{87}$Rb, $^{7}$Li, $^{41}$K (Newton-Cartan), and $^{23}$Na (Einstein-Hilbert).
\end{abstract}

\maketitle
\section{\label{sec:intro}Introduction}

The marriage of quantum mechanics and general relativity is one of the greatest outstanding problems in modern physics.
This is in part due to the fact that this theory would only become truly necessary under the most extreme conditions\textemdash the singularity of a black-hole or the initial moments after the big bang.
As such, it is extremely difficult to theoretically describe, let alone physically probe.

Despite the seeming intractability, some headway may be made in the understanding of such extreme theories by way of analogy.
This idea traces back to Unruh, who in 1981\cite{unruhExperimentalBlackHoleEvaporation1981} suggested that a flowing quantum fluid could realize a laboratory scale analog of a quantum field theory in a curved spacetime.
Access to even the most rudimentary quantum simulator for such a curved spacetime could provide valuable insights into this otherwise inaccessible regime.

Since Unruh's initial proposal, many systems have been advanced as candidates for realizing analog spacetimes~\cite{barceloAnalogueGravity2011}, including liquid helium~\cite{jacobsonEventHorizonsErgoregions1998,volovikUniverseHeliumDroplet2009,volovikTopologyQuantumVacuum2013}, Bose-Einstein condensates~\cite{garaySonicAnalogGravitational2000,*garaySonicBlackHoles2001,eckelRapidlyExpandingBoseEinstein2018,keserAnalogueStochasticGravity2018,macherBlackholeRadiationBoseEinstein2009,fischerQuantumSimulationCosmic2004,fedichevGibbonsHawkingEffectSonic2003,schutzholdSweepingSuperfluidMott2006,chaProbingScaleInvariance2017,steinhauerObservationSelfamplifyingHawking2014,*steinhauerObservationQuantumHawking2016}, nonlinear optical media~\cite{leonhardtRelativisticEffectsLight2000}, electromagnetic waveguides~\cite{schutzholdHawkingRadiationElectromagnetic2005}, magnons in spintronic devices~\cite{roldan-molinaMagnonicBlackHoles2017}, semi-conductor microcavity polaritons~\cite{nguyenAcousticBlackHole2015}, Weyl semi-metals~\cite{volovikBlackHoleHawking2016,kedem2020black}, and even in classical water waves~\cite{euveObservationNoiseCorrelated2016}.
Analog gravity systems are no longer a theoretical endeavor; recent experiments have realized the stimulated Hawking effect~\cite{droriObservationStimulatedHawking2019}, and in the case of a Bose-Einstein condensate a spontaneous Hawking effect~\cite{steinhauerObservationSelfamplifyingHawking2014,*steinhauerObservationQuantumHawking2016}.

\begin{table}[t]
    \centering
    \begin{ruledtabular}
    \begin{tabular}{cccc}
         Goldstone mode & Dispersion & Analog spacetime & Lagrangian \\
        %   mode     &            & spacetime & \\
         \hline 
         Type-I & $\omega \sim k$ & Einstein-Hilbert  & Eq.~\eqref{eq:relativistic-scalar}\\
         Type-II & $\omega \sim k^2$ & Newton-Cartan  & Eq.~\eqref{eq:NC-lagrangian-II}
    \end{tabular}
    \end{ruledtabular}
    \caption{Analog spacetimes which appear for the different Goldstone modes in the presence of a background condensate flow.
    These spacetimes emerge as effective field theories governing the long-wavelength behavior.
    As we demonstrate in this work, the emergent geometry is determined by the flow profile of the background condensate.
    This is explicitly demonstrated in Sec.~\ref{sec:relativistic} for the Type-I modes and Sec.~\ref{sec:nonrelativistic} for the Type-II modes, where we also provide an overview of the Newton-Cartan formalism.
    }
    \label{tab:Key_results}
\end{table}

In this paper we introduce an analog gravity system that exhibits Newton-Cartan geometry~\cite{cartanVarietesConnexionAffine1923,*cartanVarietesConnexionAffine1924,son2013newtoncartan}.
This geometry naturally arises from a full analysis of all Goldstone modes in a flowing spinor (or multicomponent) condensate.
Spinor condensates  \cite{stamper-kurnSpinorBoseGases2013} have been studied in the context of analog curved space before~\cite{fischerQuantumSimulationCosmic2004,weinfurtnerAnalogueSpacetimeBased2007}; however a full accounting of all gapless modes has not been done to the best of our knowledge.
The Goldstone modes which realize the Newton-Cartan geometry exhibit a quadratic $\omega\sim \mathbf{k}^2$ dispersion, known as ``Type-II'' Goldstone modes~\cite{watanabeUnifiedDescriptionNambuGoldstone2012,hidakaCountingRuleNambuGoldstone2013}.
For example, the spin wave excitations about an SU(2) symmetry breaking ferromagnetic mean-field are such a mode.
Distinct from the linearly dispersing case (called ``Type-I'' modes), Newton-Cartan spacetimes implement local Galilean invariance, as opposed to local Lorentz invariance.
These results are general and summarized in Table~\ref{tab:Key_results}, where we give a general prescription for separating out all Goldstone modes into either Type-I (linearly dispersing) or Type-II (quadratically dispersing) modes and assigning them either an Einstein-Hilbert or Newton-Cartan spacetime geometry.
{In the process of determining the analogue spacetimes of various Goldstone modes, we also generalize the existing proofs of non-relativistic Goldstone theorems~\cite{watanabeUnifiedDescriptionNambuGoldstone2012,hidakaCountingRuleNambuGoldstone2013} to allow for inhomogeneous mean-field textures. 
To this end, we explicitly show in very general terms, how the Goldstone modes couple to the spacetime variations in the mean-field texture, which in principle paves the way for the applications towards the study of nonequilibrium symmetry breaking dynamics beyond the paradigm of analog gravity.
It is of central importance to our work that the symplectic structure which distinguishes the different types of Goldstone modes still remains even in the inhomogeneous case, and this explicitly shown in our proof.}

Newton-Cartan geometry was developed by Cartan~\cite{cartanVarietesConnexionAffine1923,*cartanVarietesConnexionAffine1924} and refined by others \cite{kunzleGalileiLorentzStructures1972} as a geometric formulation and extension of Newtonian gravity.
It has since found application across different areas of physics, including in quantum Hall systems \cite{son2013newtoncartan,gromovThermalHallEffect2015,bradlynLowenergyEffectiveTheory2015} and effective theories near Lifshitz points \cite{christensenBoundaryStressenergyTensor2014,christensenTorsionalNewtonCartanGeometry2014} with interest to the high-energy community with implications for quantum gravity \cite{hartongHoravaLifshitzGravityDynamical2015,taylorLifshitzHolography2016}.
We extend these applications here to flowing condensates for the case of Type-II Goldstone modes.

Heuristically, one may view the quadratic dispersion relation $\omega \sim |\mathbf{k}|^2 + ...$ as the limit of a linear dispersion relation $\omega \sim v|\mathbf{k}| + ...$ with vanishing group velocity $v\rightarrow 0$.
In terms of the analog spacetime, this corresponds to an apparent vanishing of the speed of light.
As such, the formation of event horizons and their corresponding Hawking radiation ought to be ubiquitous in such spacetimes; however our results contradict this intuition.
Specifically, we find that fields propagating in Newton-Cartan geometries exhibit an additional conservation law which precludes the emission of Hawking radiation.
{It is worth remarking that similar constraints on magnon scattering amplitudes in an $SU(2)$-symmetric ferromagnet have been discussed in the context of nonequilibrium kinetic theories~\cite{bhattacharyyaUniversalPrethermalDynamics,rodriguez-nievaHydrodynamicSoundModes}.
It would be interesting to connect these two observations in future works.}

The immediate implication of this is that any Type-I mode can have an effective event horizon and therefore a Hawking effect (similar things have been noticed for specific other Type-I modes), and further, no Hawking effect can occur for Type-II modes, at least not without introducing quasiparticle interactions (which corresponds to going being a quadratic treatment of fluctuations).

Finally, we discuss the relationship between transport phenomena and gravitational metrics in our theory~\cite{luttingerTheoryThermalTransport1964,gromovThermalHallEffect2015,geracieSpacetimeSymmetriesQuantum2015,son2013newtoncartan}.
Specifically, we obtain the stress-tensor, energy flux, and momentum density for theories both with the Einstein-Hilbert and Newton-Cartan geometries.
In particular, we relate the energy-momentum tensor calculated in an analog Einstein-Hilbert geometry to its nonrelativistic counterparts through the use of Newton-Cartan geometry.
This helps identify how the analog Hawking effect results in nontrivial energy and momentum currents in the underlying nonrelativistic system.

The outline of the paper is as follows.
Section~\ref{sec:spacetimes} {contains the generalization of Goldstones theorem to ``curved" mean-field profiles, and} shows that in the presence of a flowing background condensate Type-I and -II Goldstone modes couple to Einstein-Hilbert (Section~\ref{sec:relativistic}) and Newton-Cartan (Section~\ref{sec:nonrelativistic}) geometries respectively.
In Section~\ref{sec:model}, we present a minimal model for these space-times and the phase transition that connects them.
In Sec.~\ref{subsec:bdg} we develop the Bogoliubov-de Gennes framework which we then use to analyze this system.
In Sec.~\ref{sec:step} we apply this to a specific step-like flow geometry and show the effect of the geometry on the emitted Hawking radiation.
We then discuss transport of energy and momentum in these different analog spacetimes systems in Sec.~\ref{sec:transport}.
We conclude the paper in Section~\ref{sec:conclusion}.
Our two appendices include Appendix~\ref{app:fluctuations} where we put the full fluctuation calculation of the Lagrangian and Appendix~\ref{app:BogoliubovHawking} where we review the Hawking calculation for the phonon problem.
Throughout, we take $\hbar=k_B=1$ and our relativistic metrics have signature ($+$ $-$ $-$ $-$).
We also indicate spatial vector with a boldface (e.g. $\mathbf{r}$), while spacetime vectors are indicated without boldface (e.g. $x = (t,\mathbf{r})$).

\section{Relationship between spacetime and Goldstone's theorem}\label{sec:spacetimes}

In this work we consider models of ultra-cold bosonic spinor quantum gases described by an $N$-component field variable $\Psi(\mathbf{r},t) = [ \Psi_1, \Psi_2, \ldots, \Psi_N ]^T$ residing in $d$ spatial dimensions (we do not make the distinction between ``spinor" and higher multiplet fields in this work).
The Lagrangian describing this system is taken to be of the general form
\begin{equation}
    \mathcal L = \tfrac{i}{2}(\Psi^\dagger \overrightarrow{\partial_t} \Psi - \Psi^\dagger \overleftarrow{\partial_t} \Psi) - \tfrac1{2m} \nabla \Psi^\dagger \cdot \nabla \Psi - V(\Psi^\dagger, \Psi), \label{eq:general-Lagrangian}
\end{equation}
where $m$ is the mass of the atoms in the gas and $V(\Psi^\dagger, \Psi)$ is a general potential energy function that includes interactions with an external potential as well as local inter-particle interactions.
Such a system may be realized by cold-atoms, where in addition to the inter-particle interactions external potentials such as a harmonic trap, optical lattice, or magnetic field may be present.
For a comprehensive review regarding the theory and experimental realization of spinor condensates see Ref.~\cite{stamper-kurnSpinorBoseGases2013}.

We consider the case where the Lagrangian exhibits invariance under an internal symmetry described by a Lie group $G$, according to which $\Psi$ transforms {under the fundamental representation (we henceforth do not distinguish between the symmetry group and its representation),} such that the action $\mathcal{S} = \int \mathcal{L}\ d^{d+1} x$ remains invariant.
That is,
\begin{equation}
    \Psi(x) \rightarrow U \Psi(x)\Rightarrow \mathcal{S} \rightarrow \mathcal{S} \quad \forall U \in G.
\end{equation}
Recall that a Lie group $G$ is generated by its corresponding Lie algebra $\mathfrak{g}$, and this has a representation of $\mathcal{R}(\mathfrak{g})$ when acting on the field $\Psi$.
For ease of calculations, we use the mathematical convention that Lie algebras consist of anti-Hermitian elements.
Hence, if $A$ is an element of $\mathcal{R}(\mathfrak{g})$, then $A = -A^\dagger$ and the corresponding group element is $e^{A} = ((e^{A})^{-1})^\dagger$.

We pursue a semi-classical analysis of our system by first obtaining the classical equations of motion (i.e. the saddle-point of the action).
Then we linearize the action around the saddle-point, obtaining a description of the symmetry-broken phases in terms of their Goldstone modes.
The primary point of our work is that this linearized action admits a simple description in terms of different emergent analog spacetimes and depending on the nature of the saddle-point, this analog spacetime may develop non-trivial curved geometry.

The rest of this section is organized as follows.
We perform a quadratic fluctuation analysis in Section~\ref{sec:fluctuations}.
In Section~\ref{sec:nonrel-goldstone-proof} we review the proof of the Goldstone theorem in non-relativistic settings~\cite{watanabeUnifiedDescriptionNambuGoldstone2012,hidakaCountingRuleNambuGoldstone2013} and show how this allows us to classify Goldstone modes into Type-I and Type-II.
Section~\ref{sec:full-lagrangian} then presents the full Lagrangian for the Goldstone modes while Sections~\ref{sec:relativistic} and~\ref{sec:nonrelativistic} make explicit the connection to curved space geometry.

\subsection{Saddle-Point Expansion}\label{sec:fluctuations}

We begin by looking for saddle-points of the Lagrangian Eq.~\eqref{eq:general-Lagrangian}, the spinor Gross-Pitaevskii equation
\begin{equation}
    i\partial_t \Psi = -\frac1{2m}\nabla^2 \Psi + \frac{\partial V}{\partial \Psi^\dagger}. \label{eq:EulerLagrange}
\end{equation}
Suppose that we have found a mean-field solution to this equation $\Psi_0(\mathbf{r},t) \equiv \braket{\Psi(\mathbf{r},t)}$ which describes the dynamics of a mean-field condensate (neglecting fluctuation back-reaction); for a general out-of-equilibrium system, the space-time dependence of $\Psi_0(\mathbf{r},t)$ may be non-trivial~\cite{barnettGeometricalApproachHydrodynamics2009,keserAnalogueStochasticGravity2018,stamper-kurnSpinorBoseGases2013}.

The presence of a non-zero mean-field solution $\Psi_0$ spontaneously breaks the internal symmetry group $G$ down to a subgroup $H \subset G$.
Let $\mathfrak{h}$ be the Lie algebra that generates the subgroup $H$.
This is defined by the set of generators
\begin{equation}
    \mathfrak{h} = \{ \bm \tau\in \mathfrak{g} \;|\; \bm \tau \Psi_0 = 0 \}.
\end{equation}
We can form a complete basis for $\mathfrak{h} = \spn\{\tau_k\}$.
The original Lie algebra then separates into two sub-spaces; $\mathfrak{g} = \mathfrak{h} \oplus \mathfrak{h}^c$, where $\mathfrak{h}^c$ is simply the complement of $\mathfrak{h}$.
It is useful to form an explicit basis for $\mathfrak{h}^c \equiv \spn\{\sigma_l\}$ so that $\mathfrak{g} = \spn\{\tau_k\}\cup\{ \sigma_l\} = \spn\{\sigma_l, \tau_k\}$.
Formally, $\mathfrak{h}^c$ is isomorphic to the quotient algebra $\mathfrak{g}/\mathfrak{h}$, and the basis elements $\sigma_l$ are isomorphic to coset spaces.

{For the sake of simplicity, in this work we will only concern ourselves with systems which have homogeneous spin orders, and focus on the effects of inhomogeneous condensate textures, as this is already very interesting and non-trivial.
However, our proof can be extended to include the most general case which has both inhomogeneous spin and condensate textures.
The resulting expression for the Goldstone mode effective action is given in Appendix~\ref{app:spin-textures}.
Studying the effects of inhomogeneous spin order is both challenging and of great interest, as it involves the introduction of a non-Abelian connection in spacetime.
We leave this problem open, to be addressed in future works. 

More precisely, we will assume that, although in general the mean-field $\Psi_0(x)$ may break the symmetry group $G$ down to different subgroups $H=H(x)$ at each spacetime point, we will only consider mean-fields which have subgroups $H(x)$ which only differ in the Abelian phase subgroup, and thus have a homogeneous spin mean-field.}  %we do not consider this in full generality since it leads to a very complicated (but interesting) structure involving a non-Abelian connection on the spacetime.
%However, we later consider {\it flowing} condensates which inhomogeneously break the $U(1)$ subgroup of $G$.

We now examine the quadratic fluctuations of the field $\Psi$ about the mean-field by expanding the Lagrangian in powers of $\delta\Psi(x) = \Psi(x) - \Psi_0(x)$.
This separates into two distinct contributions; the massless Goldstone modes $\theta_l(x)$ which correspond to spontaneously broken symmetries, and massive fields $\beta_n(x)$ which describe all the remaining modes.
Each Goldstone mode corresponds to a broken generator $\sigma_l \in \bar{\mathfrak{h}}$ acting on the mean-field condensate $\Psi_0(x)$.
These contribute to the fluctuation action as
\begin{equation}
\left(\delta \Psi(x) \right)_{\textrm{Goldstone}} = \sum_{l} \theta_l(x) \sigma_l \Psi_0(x)\equiv \bm\sigma(x) \Psi_0(x) ,\label{eq:goldstone-modes-define}
\end{equation}
which serves to define the Goldstone matrix field $\bm\sigma(x)$.
The remaining degrees of freedom are generically massive and are not amenable to a description in terms of the Lie algebra's generators.
It is advantageous to parameterize the fluctuations $\delta \Psi$ in terms of real fields with massive terms orthogonal to the massless terms in the sense described below.
Within the quadratic theory, this implies the fluctuations reside within a real vector space $\mathbb{R}^{2N}\sim \mathbb{C}^N$.
The Goldstone modes $\sigma_l \Psi_0(x)$ form a subspace of this manifold while the remaining basis elements are generically massive and are written as $\xi_n(x)$.
We note that in general the basis elements are spacetime dependent simply because the mean-field is also spacetime dependent.

In order to make the notion of orthogonality precise we lift the standard complex ($\mathbb{C}^N$) inner product onto our real vector space $\mathbb{R}^{2N}$ to obtain the \emph{real} inner product $g$ defined by
\begin{equation}
    g(\xi,\chi) \equiv \tfrac12 (\xi^\dagger \chi + \chi^\dagger \xi) \label{eq:real-inner-product}.
\end{equation}
In terms of the Goldstone manifold and its complement, the variation $\delta \Psi(x) $ takes the compact form
\begin{equation}
    \delta \Psi(x) = \bm\sigma(x)\Psi_0(x) + \xi(x) ,\label{eq:fluctuation-expanded}
\end{equation}
where we have defined the massive modes by
\begin{equation}
    \xi(x) = \sum_n \beta_n(x) \xi_n(x) .\label{eq:massive-modes-define}
\end{equation}

We proceed to the expansion of the Lagrangian in terms of the variation $\delta \Psi$.
First, we consider the potential.
It is locally invariant under under $G$, so we can write
\begin{equation}
  V(\Psi^\dagger, \Psi) = V(\Psi^\dagger e^{\bm \sigma(x)}, e^{-\bm\sigma(x)} \Psi).
\end{equation}
Furthermore, we can use our expansion of $\Psi(x)$ to obtain
\begin{equation}
\begin{split}
    e^{-\bm\sigma} \Psi & \approx e^{-\bm\sigma}[\Psi_0 + \bm\sigma\Psi_0 + \xi] \\
     & \approx (1 - \bm\sigma +\tfrac12 \bm\sigma^2)[\Psi_0 + \bm\sigma\Psi_0 + \xi] \\
     & \approx \Psi_0 + \xi - \bm \sigma \xi  - \tfrac12 \bm\sigma^2 \Psi_0,
\end{split}
\end{equation}
keeping terms up to quadratic order in fluctuations.
This allows us to expand the potential energy up to quadratic order (dropping the terms constant and linear in the variation)
\begin{multline}
  V(\Psi^\dagger, \Psi) = - \left[ \frac{\partial V}{\partial \Psi}\cdot \left(\tfrac12 \bm \sigma^2 \Psi_0 + \bm  \sigma \xi \right) + \mathrm{c.c.} \right]\\
   +\frac12 \xi^* \xi^*\cdot \frac{\partial^2 V}{\partial \Psi^\dagger \partial \Psi^\dagger} +  \xi^*\cdot \frac{\partial^2 V}{\partial \Psi^\dagger \partial \Psi} \cdot \xi + \frac12 \frac{\partial^2 V}{\partial \Psi \partial \Psi}\cdot \xi \xi , \label{eq:potential-expand}
\end{multline}
where all derivatives of the potential are understood as being evaluated at the mean-field.
The terms quadratic in $\xi,\xi^*$ represent massive terms, and the first line of Eq.~\eqref{eq:potential-expand} drops out when combined on-shell with similar terms from the kinetic part of the Lagrangian.
Deriving the full fluctuation Lagrangian is not instructive, and has been relegated to Appendix~\ref{app:fluctuations}; the final result is given below.

Focusing on the Goldstone modes, written in terms of the ``angle fields" $\theta_l(x)$, the resulting Lagrangian for fluctuations is given by
\begin{multline}
  \mathcal L_{\mathrm{fluc}} = \theta_m P^{\mu}_{mn} (\partial_\mu \theta_n)
    + \beta_m Q^{\mu}_{mn} (\partial_\mu \theta_n) \\ + (\partial_j\theta_n)T^{jk}_{mn}(\partial_k \theta_n) + \mathcal L_{\mathrm{mass}}(\beta_m, \partial_\mu \beta_m), \label{eq:partial-fluctuation-Lagrangian}
\end{multline}
where we have instituted the Einstein summation convention.
In this and the following, Roman indices $i,j,k,\ldots$ run over spatial dimensions while Greek indices $\mu,\nu,\ldots$ run over both temporal and spatial dimensions (with $\mu = 0 = t$ the temporal index).
The Roman indices $n,m,\ldots$ enumerate the different Goldstone modes or massive modes and are similarly summed.
The terms $P^{\mu}_{mn}$, $Q^{\mu}_{mn}$, and $T^{jk}_{mn}$ depend on both space and time, and are given by
\begin{equation}
  \begin{split}
  P^{t}_{mn} & = \tfrac{i}{2} \Psi_0^\dagger [\sigma_n, \sigma_m] \Psi_0, \\
  P^{j}_{mn} & = \tfrac1{4m} ( \partial_j \Psi_0^\dagger [\sigma_m, \sigma_n] \Psi_0 - \Psi_0^\dagger [\sigma_m, \sigma_n] \partial_j \Psi_0), \\
Q^{t}_{mn} & = i (\Psi_0^\dagger \sigma_n \xi_m + \xi_m^\dagger \sigma_n \Psi_0), \\
Q^{j}_{mn} & = \tfrac1{2m} ( \xi_m^\dagger \sigma_n \partial_j \Psi_0 - \partial_j \Psi_0^\dagger \sigma_n \xi_m \\ & \phantom{= = \quad \quad \quad\quad } + \Psi_0^\dagger \sigma_n \partial_j \xi_m - \partial_j \xi_m^\dagger \sigma_n \Psi_0), \\
T^{jk}_{mn} & = \tfrac{1}{2m} \delta^{jk} \Psi_0^\dagger \sigma_n \sigma_m \Psi_0.
\end{split}
\end{equation}
As mentioned previously, it is also important to keep track of the massive modes in the full Lagrangian and we offer that full analysis in Appendix~\ref{app:fluctuations}.

\subsection{Proof of the nonrelativistic Goldstone theorem}\label{sec:nonrel-goldstone-proof}

Before proceeding to simplify the Lagrangian and derive the curved space analogues, we need to understand and make use of the nonrelativistic Goldstone theorem~\cite{hidakaCountingRuleNambuGoldstone2013,watanabeUnifiedDescriptionNambuGoldstone2012}, providing a complementary proof in the process.

We consider the following ansatz for the mean-field
\begin{equation}
  \Psi_0(x) = \sqrt{\rho(x)} e^{i\vartheta(x)} \chi, \quad \chi^\dagger \chi = 1, \quad \partial_\mu \chi = 0. \label{eq:mean-field_ansatz}
\end{equation}
Importantly the spinor structure given by $\chi$ is independent of space and time.
{ This ansatz is justified since if we prepare a static mean-field state and quench it dynamically with terms that only couple to mass (such that the Gross-Pitaevskii equation describes its dynamics), there are no terms in the hydrodynamics that generate spin texture (i.e., the dynamics will preserve $\partial_\mu \chi_\mu=0$) \cite{stamper-kurnSpinorBoseGases2013,lamacraftLongwavelengthSpinDynamics2008,barnettGeometricalApproachHydrodynamics2009}.}
The global $U(1)$ symmetry implies the phase and density obey a continuity relation which can be conveniently written as
\begin{equation}
    \partial_\mu J^\mu = 0 ,
\end{equation}
with the condensate four-current given by $J^\mu = \rho v_s^\mu$, where the superfluid four-velocity field is $v_s^\mu = (1, \frac{1}{m}\nabla \vartheta )$.
This simplifies the term
\begin{equation}
  P^\mu_{mn} = -\tfrac{i}2 J^\mu \chi^\dagger [\sigma_n,\sigma_m] \chi,
\end{equation}
which dictates which real fields $\theta_n$ are canonically conjugate to each other.
In non-relativistic systems, the relationship between broken symmetry generators and Goldstone modes is not one-to-one.
Instead, we must separate out our modes into Type-I and Type-II Goldstone modes, which is done by going to the preferred basis of the matrix $P^\mu_{mn}$.

To understand this, we return to the \emph{real} vector space defined by the Goldstone mode manifold, which we label $\mathcal{A}_{\mathbb{R}}$.
That is,
\begin{equation}
    \mathcal{A}_{\mathbb{R}} = \spn_{\mathbb{R}}\{\sigma_l \Psi_0(x) \}.\label{eq:real-Goldstone}
\end{equation}
The real dimension $D_\mathbb{R}$ of this subspace is simply equal to the number of broken generators.
We can complexify this vector space by allowing for complex-valued coefficients
\begin{equation}
  \mathcal{A}_{\mathbb{C}} \equiv \spn_{\mathbb C}\{ \sigma_n \Psi_0 \}.
\end{equation}
It may be the case that two generators which are linearly independent under real coefficients are linearly dependent when multiplied by complex coefficients.
For this reason, this vector space has an associated \emph{complex} dimension $D_{\mathbb{C}} \leq D_{\mathbb{R}}$.
The essence of the Goldstone mode theorem is that $D_{\mathbb{R}}$ is the number of broken generators and $D_{\mathbb{C}}$ is the number of modes, and these two quantities can be formally related by classifying each basis element $\sigma_l \Psi_0(x) \in \mathcal{A}_{\mathbb{R}}$ due to whether $i\sigma_n \Psi_0 \in \mathcal{A}_{\mathbb{R}}$ or not.

To establish this we need to return to our real inner product $g(\cdot,\cdot)$.
We can use the operation of multiplication by $i$ to define a symplectic bilinear form $\omega(\cdot,\cdot)$ by
\begin{equation}
  \omega(\eta, \xi) \equiv g(i\eta, \xi) = \tfrac{i}2(\xi^\dagger \eta - \eta^\dagger \xi).
\end{equation}
The multiplication by $i$ (acting on the basis vectors $\sigma_l \Psi_0(x)$) can be restricted to the real vector space $\mathcal{A}_{\mathbb{R}}$, which we define by the notation
\begin{equation}
   i|_{\mathcal{A}_{\mathbb{R}}} \equiv I : \mathcal{A}_{\mathbb{R}} \rightarrow \mathcal{A}_{\mathbb{R}}.
\end{equation}
Similarly, we define $\rng I \equiv \mathcal{A}_{\mathrm{II}} \subset \mathcal{A}_{\mathbb{R}}$ as the range of $I$.
The null space of $I$ is then defined to be $\mathcal A_{\mathrm{I}}$ and represents states $\eta \in \mathcal{A}_{\mathbb{R}}$ which leave the real vector space upon multiplication by $i$.
As a simple example, consider unit vectors $\hat{e}_1 = (1,0)^T$ and $\hat{e}_2 = (i,0)^T$.
As elements of a real vector space these are linearly independent, however $i \hat{e}_1 = \hat{e}_2$ and so these are not linearly independent in a complex vector space.
In this case, we have $D_{\mathbb{R}} = 2,\ D_{\mathbb{C}} = 1$ and $\rng I = \mathcal{A}_{\mathbb{R}},\ \nll I = 0$.
However, if $\hat{e}_1 = (1,0)^T$ and $\hat{e}_2 = (0,1)^T$ then $D_{\mathbb{R}} = 2 = D_{\mathbb{C}}$ and $\rng I = 0,\ \nll I = \mathcal{A}_{\mathbb{R}}$.

The classification of basis elements may be accomplished by taking the real inner product of $i\eta$ with the other elements of $\mathcal A$\textemdash if this vanishes, then $\eta$ is in the kernel of $I$.
But this is exactly given by the symplectic bilinear form defined above so that
\begin{equation}
  \mathcal{A}_{\mathrm{I}} \equiv \nll I = \{ \eta\in \mathcal{A}_{\mathbb{R}} \;|\;\omega(\eta, \chi) = 0, \forall \chi \in \mathcal A\}.
\end{equation}
This condition can be simplified into a matrix condition if we note that we can let $\eta = \sum_n a_n \sigma_n \Psi_0$ and $\chi = \sum_m b_m \sigma_m \Psi_0$, so that
\begin{equation}
  0 = \omega(\eta,\chi) = -\tfrac{i}2 a_n \Psi_0^\dagger [\sigma_n,\sigma_m] \Psi_0 b_m.
\end{equation}
This relates the null-space of $I$ to the null-space of the matrix $\Psi_0^\dagger [\sigma_n,\sigma_m] \Psi_0 \propto P^\mu_{mn}$, the term appearing in our Lagrangian which determines the canonically conjugate pairs of modes.
Using the rank-nullity theorem, we have
\begin{equation}
  \mathcal{A}_{\mathbb{R}} = \mathcal{A}_{\mathrm I} \oplus \mathcal{A}_{\mathrm{II}}.
\end{equation}
Since the matrix given by elements $-\tfrac{i}2\Psi_0^\dagger [\sigma_n,\sigma_m] \Psi_0$ is real and antisymmetric, we can block-diagonalize the matrix with a special orthogonal transformation.
Going to this basis and using our ansatz for the flowing mean-field $\Psi_0 = \sqrt{\rho} e^{i\vartheta} \chi$, the result is
\begin{equation}
  \begin{split}
  -\tfrac{i}2\Psi_0^\dagger& [\sigma_n,\sigma_m] \Psi_0  =  -\tfrac{i}2  \rho \chi^\dagger [\sigma_n,\sigma_m] \chi \\ \phantom{\Big(} \\
& = \rho
\begin{pmatrix}[ccccc|cc]
  \tikzmark{x1} 0 & \lambda_1 & 0          & 0         & \phantom{0}\tikzmark{x2}       & \tikzmark{y1}\phantom{0} & \phantom{0}\tikzmark{y2}                \\
  -\lambda_1 & 0         & 0          & 0         & \cdots & 0     & \cdots  \\
  0          & 0         & 0          & \lambda_2 &        &                 \\
  0          & 0         & -\lambda_2 & 0         &        &                 \\
             & \vdots    &            &           & \ddots &                 \\
  \hline
             & 0         &            &           &        & 0      &        \\
             & \vdots    &            &           &        &        & \ddots
\end{pmatrix},
\end{split}\label{eq:typeiandtypeiimatrix}%
\begin{tikzpicture}[overlay, remember picture,decoration={brace,amplitude=2pt}]
\draw[decorate,thick] (x1.north) -- (x2.north)
      node [midway,above=5pt] {$\mathcal A_{\mathrm{II}}$};
\draw[decorate,thick] (y1.north) -- (y2.north)
      node [midway,above=5pt] {$\mathcal A_{\mathrm{I}}$};
\end{tikzpicture}%
\end{equation}%
with $\lambda_j>0$.
This defines a preferred basis for the broken generators $\{\sigma_l\}$ which we henceforth assume is the basis we are in.
Note that in this basis $\mathcal{A}_{\mathrm{II}}$ takes the form of a direct sum of decoupled symplectic forms.

This matrix provides a natural way to break up the generators.
First, we can define $\sigma_n^{\mathrm{II}}$  and its conjugate generator  $\overline{\sigma_n^{\mathrm{II}}}$ via $-\tfrac{i}2 \Psi_0^\dagger[\sigma_n^{\mathrm{II}},\overline{\sigma_n^{\mathrm{II}}}]\Psi_0 = \rho \lambda_n$.
This implies that $\overline{\sigma_n^{\mathrm{II}}}\Psi_0 = i\sigma_n^{\mathrm{II}}\Psi_0$ (however $\overline{\sigma_n^{\mathrm{II}}} \neq i\sigma_n^{\mathrm{II}}$).
Let $n_{\mathrm{II}}$ be the number of $\lambda_j$'s, so that $\dim(\mathcal A_{\mathrm{II}}) = 2n_{\mathrm{II}}$.
As the coefficient of the temporal derivative term in the Lagrangian, this matrix tells us that the two Goldstone fields described by $\sigma_n^{\mathrm{II}} \Psi_0(x)$ and $\overline{\sigma_n^{\mathrm{II}} }\Psi_0(x)$ are canonically conjugate to each other and therefore describe the {\bf same mode}, a Type-II Goldstone mode.
Finally, let $\dim(\mathcal A_{\mathrm{I}}) = n_{\mathrm{I}}$ be dimension of the null-space of $I$.
This is the number of Type-I Goldstone modes; they represent modes which are canonically conjugate to a massive mode.
It is evident by the rank-nullity result that
\begin{equation}
    2n_{\mathrm{II}} + n_{\mathrm{I}} = D_{\mathbb{R}}
\end{equation}
is the number of broken generators, while
\begin{equation}
    n_{\mathrm{II}} + n_{\mathrm{I}} = D_{\mathbb{C}}
\end{equation}
is the number of Goldstone modes in the system.

With this particular grading into $n_{\mathrm{II}}$ basis elements $\sigma_n^{\mathrm{II}}\Psi_0$ and $n_{\mathrm{I}}$ basis elements $\sigma_n^{\mathrm{I}}\Psi_0$, we can rewrite our real vector space
\begin{equation}
  \mathcal{A}_{\mathbb{R}} = \spn \{ \sigma_n^{\mathrm{II}}\Psi_0, \overline{\sigma_n^{\mathrm{II}}}\Psi_0, \sigma_n^{\mathrm{I}}\Psi_0 \},
\end{equation}
and similarly, we can write the complexified vector space in two equivalent ways
\begin{equation}
  \begin{split}
  \mathcal{A}_{\mathbb{C}} & = \spn_{\mathbb{C}} \{ \sigma_n^{\mathrm{II}}\Psi_0, \sigma_n^{\mathrm{I}}\Psi_0 \}, \\
  \mathcal{A}_{\mathbb{C}} & = \spn \{ \sigma_n^{\mathrm{II}}\Psi_0, \overline{\sigma_n^{\mathrm{II}}}\Psi_0, \sigma_n^{\mathrm{I}}\Psi_0, i \sigma_n^{\mathrm{I}}\Psi_0 \}.
\end{split}
\end{equation}
The modes represented by $i \sigma_n^{\mathrm{I}}\Psi_0$ are exactly the massive modes conjugate to $\sigma_n^{\mathrm{I}}\Psi_0$ (by definition, they are not in $\mathcal A$ and are thus not associated with a broken generator).

{ In fact, as we have shown $P^t_{mn} = - \omega(\sigma_n \Psi_0, \sigma_m \Psi_0)$ while for the massive modes $Q^t_{mn} = -2\omega(\sigma_n\Psi_0,\xi_m)$.
These two matrices have different images, as we can see since by construction $\omega(\sigma_n^\mathrm{I}\Psi_0,\sigma_m\Psi_0)=0$ while $\omega(\sigma_n^\mathrm{II}\Psi_0,\xi_m)=g(\overline{\sigma_n^\mathrm{II}} \Psi_0,\xi_m)=0$. 
In other words, $P^t$ has range $\mathcal A_\mathrm{II}$, and  $Q^t$ has range $\mathcal A_\mathrm{I}$ due to the real fields being orthogonal by Eq.~\eqref{eq:real-inner-product}.}

At low energies (below the relevant mass gaps), massive modes that are not conjugate to Goldstone modes can be trivially integrated out and do not contribute in the IR.
This then leaves the Goldstone modes, which are gapless, and a few massive modes which are canonically conjugate to the Type-I Goldstone modes.
These massive modes cannot be trivially integrated out and they are to be included in the low-energy theory.
Doing so amounts to adding the basis elements $i \sigma_n^{\mathrm{I}} \Psi_0$ to our fluctuation manifold.

\subsection{Lagrangian for Goldstone Modes}\label{sec:full-lagrangian}

We now employ this classification into Type-I and -II modes to our benefit by using it to simplify the fluctuation Lagrangian.
Recall that in this work we restrict ourselves to flowing condensates which have a spatial texture to the phase mode (and thus inhomogeneously break the global $U(1)$ part of the symmetry group), but have a homogeneous and static spinor texture.
For instance, one may consider a condensate of pseudo-spin-$\frac12$ atoms in its ferromagnetic phase which has a definite homogeneous magnetization $\braket{S_z} = \chi^\dagger S_z \chi = \frac12$ but a non-zero density and phase profile.
As remarked earlier, this flow produces a non-zero spatial component for the Noether current $J_\mu(x)$.
Going to the preferred basis of $P^\mu_{mn}$, obtained in Sec.~\ref{sec:nonrel-goldstone-proof} then yields the partitioning into the Goldstone modes given by $\{ \sigma_n^{\mathrm{II}}\Psi_0, \overline{\sigma_n^{\mathrm{II}}}\Psi_0, \sigma_n^{\mathrm{I}}\Psi_0 \}$.
Let us remind the reader that Type-I modes are those for which $i\sigma_n \Psi_0$ cannot be written as a broken generator $\sigma'_n\Psi_0$ and therefore, the associated real field comes with a massive term in the Lagrangian.

The basis elements $\{ \sigma_n^{\mathrm{II}}\Psi_0, \sigma_n^{\mathrm{I}}\Psi_0 \}$ have the property that they are orthogonal in the conventional sense (e.g.\ $\eta^\dagger \chi = 0$).
As a result of this,
\begin{equation}
  \begin{split}
  \Psi_0^\dagger \sigma_n^{\mathrm{I}} \sigma_m^{\mathrm{II}} \Psi_0  & = 0, \\
  -\Psi_0^\dagger \sigma_n^{\mathrm{II}} \sigma_m^{\mathrm{II}} \Psi_0 & = \lambda_n \delta_{nm} \rho(x) , \\
  -\Psi_0^\dagger \sigma_n^{\mathrm{I}} \sigma_m^{\mathrm{I}} \Psi_0 & = \mu_n\delta_{nm} \rho(x) ,
\end{split}
\label{eq:generator_norms}
\end{equation}
where we have defined $\mu_n \equiv -\chi^\dagger (\sigma_n^{\mathrm{I}})^2 \chi>0$ and used the fact that $\lambda_n = -\chi^\dagger (\sigma_n^{\mathrm{II}})^2 \chi>0$.

In this basis, the field variation $\delta\Psi(x)$  may be described by three real Goldstone fields $\theta_n$, $\bar \theta_n$, and $\phi_n$ along with the real massive field $\beta_n$ via
\begin{equation}
\begin{split}
    \bm\sigma & = \sum_{n=1}^{n_{\mathrm{II}}}\left( \theta_n \sigma_n^{\mathrm{II}} + \bar \theta_n  \overline{\sigma_n^{\mathrm{II}}} \right) + \sum_{n=1}^{n_\mathrm{I}} \phi_n  \sigma_n^{\mathrm{I}}, \\
    \xi & = \sum_{n=1}^{n_\mathrm{I}} \beta_n i \sigma_n^{\mathrm{I}} \Psi_0 + \cdots,
\end{split}
\end{equation}
where ``$\cdots$'' represents other massive modes that can be trivially integrated out.
In this basis, the coefficient $P^\mu_{mn}$ simplifies to
\begin{equation}
  P^{\mu}_{mn} = \delta_{n\bar{m}} \lambda_n \rho(x) v_s^\mu,
\end{equation}
where $\bar{m}$ is defined as the index of the conjugate field to the field labeled by $m$.
Similarly, we may simplify $Q^\mu_{mn}$ which connects Type-I Goldstone modes to their conjugate massive fields.
We indeed find
\begin{equation}
  Q^\mu_{mn} = 2 \delta_{nm} \mu_n \rho(x) v_s^\mu,
\end{equation}
where the massive field with index $m$ is indicated by the basis element $i\sigma_m^{\mathrm{I}}\Psi_0$.
Lastly, we have the kinetic energy term which we can separate out into its contribution to Type-I and Type-II fields
\begin{equation}
  \begin{split}
    T^{jk}_{mn}|_{\mathrm{I}} & = -\tfrac1{2m} \delta^{jk}\rho(x) \mu_n \delta_{mn} \\
    T^{jk}_{mn}|_{\mathrm{II}} & = -\tfrac1{2m} \delta^{jk}\rho(x) \lambda_n \delta_{mn}
  \end{split}
\end{equation}

{
Notice that $\lambda_n$ or $\mu_n$ can be absorbed into a redefinition of the oscillator strength of the field it corresponds to.
Therefore, we can simply absorb the $\lambda$'s into a redefinition of the Type-II modes $\theta_n,\bar{\theta}_n$, and absorb the $\mu$'s into a redefinition of the Type-I modes $\phi_n,\beta_n$. 
In principle, this would effect the coupling to external source fields, and in the case of the Type-I modes also factors into determining the speed of sound, but we are not concerned with these effects here.}
Then, substituting the form of our fluctuations, the Lagrangian is
\begin{widetext}
\begin{multline}
  \mathcal L_{\mathrm{fluc}} = \sum_{n=1}^{n_{\mathrm{I}}} \rho(x) \left[- 2\beta_n v_s^\mu(x) \partial_\mu \phi_n - \tfrac1{2m}[(\nabla \phi_n)^2 + (\nabla \beta_n)^2] - 2 m c_n^2(x) \beta_n^2 \right] \\ + \sum_{n=1}^{n_{\mathrm{II}}} \rho(x) \left\{ -v_s^\mu(x) (\bar{\theta}_n \overrightarrow{\partial_\mu} \theta_n - \bar{\theta}_n \overleftarrow{\partial_\mu} \theta_n) - \tfrac1{2m}[(\nabla \theta_n)^2 + (\nabla \bar\theta_n)^2]\right\}. \label{eq:full-fluc-Lagrangian}
\end{multline}
\end{widetext}
Since the basis for Type-I modes is not uniquely fixed by the canonical conjugate structure of Eq.~\eqref{eq:typeiandtypeiimatrix}, this leaves us free to diagonalize the mass tensor produced by the variation of the potential in Eq.~\eqref{eq:potential-expand}.
Doing so produces the effective chemical potential terms, $m c_n^2(x)$.

We end this section with a note about the validity of this fluctuation Lagrangian: it can be seen that the overall size of this action is set by the condensate density $\rho(x)$, which uniformly multiplies all terms.
Thus, the condensate density $\rho(x)$ acts to enforce the saddle-point in the sense that if it is large, the fluctuation contribution from $\mathcal{L}_{\textrm{fluc}}$ is suppressed.
This tells us that our approach ought not be valid if either the condensate density is strongly fluctuating or vanishing all-together, as might happen at finite temperatures or near e.g. the core of a vortex.
Additionally, there may be breakdowns in smaller dimensional systems, where long-range order is prohibited by Mermin-Wagner~\cite{merminAbsenceFerromagnetismAntiferromagnetism1966,hohenbergExistenceLongRangeOrder1967,colemanThereAreNo1973}.
Barring these considerations, we proceed on to study the properties of the effective field theory described in Eq.~\eqref{eq:full-fluc-Lagrangian}.
We first consider the case where the Goldstone mode is Type-I, and then we study the case of a Type-II mode.

\subsection{Type-I Goldstones: Relativistic Spacetime}\label{sec:relativistic}
Consider an isolated Type-I Goldstone mode, with Lagrangian
\begin{multline}
  \mathcal L_{\mathrm{I}} = \rho(x) [ - 2 \beta v_s^\mu(x) \partial_\mu \phi - \tfrac1{2m}[(\nabla \phi)^2 + (\nabla \beta)^2]\\ - 2 m c^2(x) \beta^2 ],
\end{multline}
we assume that $m c^2(x)$ is large enough to dominate over the kinetic energy for $\beta$, so that $\beta$ can be easily integrated out via $ m c^2(x)\beta = -2v_s^\mu \partial_\mu \phi$.
We get the resulting Lagrangian, valid at long wavelengths and times
\begin{equation}
\label{eqn:type-i-fluc}
  \mathcal L_{\mathrm{I}}^{\mathrm{eff}} = \frac{\rho(x)}{2m} \left[ \left( \frac{v_s^\mu(x)\partial_\mu \phi}{c(x)}\right)^2 - (\nabla \phi)^2\right].
\end{equation}
This describes a scalar field propagating along geodesics of an emergent space-time metric $\mathcal{G}_{\mu\nu}$ with
\begin{equation}
  \mathcal L_{\mathrm{I}}^{\mathrm{eff}} = \tfrac12\sqrt{-\mathcal G} \mathcal G^{\mu \nu} \partial_\mu \phi \partial_\nu \phi, \label{eq:relativistic-scalar}
\end{equation}
and $\mathcal G^{\mu \nu}$ given by the line-element
\begin{equation}
  ds^2 = \frac{\rho}{c} [ c^2 dt^2 - (d\mathbf x - \mathbf v dt)^2] = \mathcal{G}_{\mu \nu} dx^\mu dx^\nu.
\end{equation}
This was first observed by Unruh in Ref.~\cite{unruhExperimentalBlackHoleEvaporation1981} where he showed that metrics of the form given above can possess non-trivial features including event-horizons.
Indeed, the metric for a Schwarschild black hole can take a very similar form in certain coordinate systems.
One of the central results of this paper is the extension of this analog to include the Type-II modes, which do not have emergent Lorentz invariance.
This is shown below.

\subsection{Type-II Goldstones: Non-relativistic Spacetime}\label{sec:nonrelativistic}

We focus on a single Type-II Goldstone mode, for which there is no massive field to integrate out.
We are left with the fluctuation Lagrangian
\begin{multline}
  \mathcal L_{\mathrm{II}} = \rho(x) \{ -v_s^\mu(x) (\bar{\theta} \overrightarrow{\partial_\mu} \theta - \bar{\theta} \overleftarrow{\partial_\mu} \theta)\\ - \tfrac1{2m}[(\nabla \theta)^2 + (\nabla \bar\theta)^2]\}.
\end{multline}
To simplify things, we group the two real fields into one complex field
\begin{equation}
  \psi = \theta + i \bar{\theta},
\end{equation}
so that we have
\begin{equation}
  \mathcal L_{\mathrm{II}} =  \rho [\tfrac{i}2 v_s^\mu (\psi^* \overrightarrow{\partial_\mu} \psi - \psi^* \overleftarrow{\partial_\mu} \psi) - \tfrac{1}{2m} | \nabla \psi|^2 ]. \label{eq:single-typeii}
\end{equation}

It turns out this too has a simple geometric description in terms of an emergent curved space-time.
However, instead of being an ``Einsteinian" geometry, the resulting description is in terms of a Newton-Cartan geometry~\cite{cartanVarietesConnexionAffine1923,*cartanVarietesConnexionAffine1924,son2013newtoncartan,gromovThermalHallEffect2015,geracieSpacetimeSymmetriesQuantum2015,bradlynLowenergyEffectiveTheory2015}.

Newton-Cartan geometry consists of three key objects: $(n_\mu, v^\mu, h^{\mu \nu})$.
These are not all independent, but rather must satisfy the constraints
\begin{equation}
  n_\mu v^\mu = 1, \quad n_\mu h^{\mu\nu} = 0.
\end{equation}
Also note that the indices on these objects are given as covariant and contravariant specifically and cannot be freely raised/lowered without the definition of a metric tensor (which we describe how to construct in Sec.~\ref{sec:transport}).

To understand the geometry these objects encode, we begin with the fundamental object that enforces time's special status within a nonrelativistic theory: $n_\mu$.
As a one-form, $n_\mu$ (colloquially, we call it the ``clock'' one-form) can be imagined as a series of surfaces (foliations), and when a spacetime displacement vector is contracted with it, it gives the elapsed time in a covariant manner.
In conjunction with the clock one-form, we have the velocity field $v^\mu$, which must go forward a unit of time (hence the constraint $n_\mu v^\mu = 1$) as a four-velocity; flow along $v^\mu$ causally connects spatial surfaces.
Lastly, the spatial metric $h^{\mu\nu}$ is degenerate ($n_\mu h^{\mu\nu}=0$) since it solely describes the geometry confined to the $d$-dimensional spatial foliations.
While in what follows we describe $h^{\mu\nu}$ emerging from intrinsic properties of the fluid flow, it can also inherit extrinsic contributions (i.e.\ if the fluid is flowing on an actual curved manifold).

In the presence of this curved Newton-Cartan geometry, the Lagrangian for a massless scalar field takes the form
\begin{equation}
  \mathcal L = n_0 \sqrt{h} [\tfrac{i}2 v^\mu (\psi^* \overrightarrow{\partial_\mu} \psi - \psi^* \overleftarrow{\partial_\mu} \psi) - \tfrac{h^{\mu\nu}}{2m} \partial_\mu \psi^* \partial_\nu \psi]
  \label{eq:NC-lagrangian-II}
\end{equation}
where $h = (|\det h^{ij}|)^{-1}$\footnote{ {In this expression, we use the spatial indices $i,j$ in the expression $\det h^{ij}$ in order to emphasize the fact that this determinant is meant to be evaluated only for the projection of the metric onto the non-degenerate subspace.
In the case considered, this is equivalent to evaluating the sub-determinant of the spatial block of the metric. }} { is the determinant of the metric projected onto the non-degenerate subspace}.

The Lagrangian of a Type-II Goldstone mode may be brought into this form.
Relating Eq.~\eqref{eq:single-typeii} to Eq.~\eqref{eq:NC-lagrangian-II}, we can extract  the geometric objects $n_\mu$, $v^\mu$, and $h^{\mu\nu}$.
We see that in our systems $h^{00} = 0 = h^{0i}$, and that $h^{ij} = h^{-1/d} \delta^{ij}$ in $d$ spatial dimensions.
Therefore, we know $n_i = 0$; hence, $n_0 v^0 = 1$.
Relating terms, we have
\begin{equation}
  \begin{split}
  \sqrt{h} &  =  \rho, \\
  n_0 \sqrt{h} v^i & = \rho v_{\mathrm{s}}^i, \\
  n_0 h^{(d-2)/(2d)} & = \rho.
\end{split}
\end{equation}
This gives us the geometric quantities
\begin{equation}
  h = \rho^2 , \quad n_0 = \rho^{2/d},
\end{equation}
and hence
\begin{equation}
\label{eqn:geometry-result}
  \begin{split}
    n_\mu & = [\rho^{2/d}, \bm{0}], \\
    v^\mu & = \rho^{-2/d} v_s^\mu, \\
    h^{ij} & = \rho^{-2/d} \delta^{ij}.
  \end{split}
\end{equation}

One important aspect of Newton-Cartan geometry is the notion of ``torsion"~\cite{bergshoeffNewtonCartanGravityTorsion2017}.
Regarded as a differential form, the clock one-form $n= n_\mu dx^\mu$ is in general not an exact differential.
This is seen by taking the exterior derivative, which defines the ``torsion tensor" $\omega = dn $.
Explicitly,
\begin{equation}
    \omega_{\mu\nu} = \partial_\mu n_\nu - \partial_\nu n_\mu.
\end{equation}
It is straightforward to see that in general, the torsion tensor in our geometry is non-zero;
\begin{equation}
  \omega_{0 j} = \partial_j n_0 = \partial_j \rho^{2/d}.
\end{equation}
Were the torsion zero, we could define an absolute time coordinate $T$, from which we would get the clock one-form as $n = dT$.
While the non-zero torsion implies there is no such absolute time, we may confirm that the more general condition
\begin{equation}
    n \wedge dn = 0
\end{equation}
is satisfied.
This is a necessary and sufficient condition for the foliation of spacetime into ``space-like" sheets which are orthogonal to the flow of time~\cite{bergshoeffNewtonCartanGravityTorsion2017}.
As such, there is still a notion of causality in this geometry.

We conclude by commenting that the Newton-Cartan geometry we find here is in fact intimately related to the gravitational field first considered by Luttinger in the context of calculating heat transport~\cite{luttingerTheoryThermalTransport1964}.
In that limit $n_\mu \propto [e^\Phi, \bm 0]$, and so the gravitational potential (up to scale factor in the logarithm) would be
\begin{equation}
  \Phi = \frac{2}d \log(\rho).
\end{equation}
Using this connection, quantities like energy current and the stress-momentum tensor can be calculated as we discuss in Sec.~\ref{sec:transport}.
First, we explore a minimal realization of these geometries and the associated quantum phases in Sec.~\ref{sec:model} as well as the fate of the Hawking effect across such a transition in Sec.~\ref{sec:step}.

\section{Minimal Theoretical Model}\label{sec:model}

In this section, we introduce a minimal model which exhibits a transition between an Einstein-Hilbert and Newton-Cartan spacetime.
We begin by analyzing the ground state within mean-field theory.
Once this is understood, we study the behavior of fluctuations about the mean-field by employing a Bogoliubov-de Gennes (BdG) description.

The model is that of a pseudo-spin-$\frac12$ bosonic field $\Psi(x) = \left(\Psi_{\uparrow}(x), \Psi_{\downarrow}(x)\right)^T$ with the following Lagrangian density
\begin{multline}
    \label{eqn:lagrangian}
    \mathcal{L} = \Psi^\dagger\left(i\partial_t + \frac{1}{2m}\nabla^2 + \mu \right)\Psi -\frac12g_0\left(\Psi^\dagger \Psi\right)^2 \\ -\frac12g_3\left(\Psi^\dagger \sigma_3 \Psi\right)^2
\end{multline}
where $\sigma_j$ are the Pauli matrices for the pseudo-spin and $\mu$ is the chemical potential, which controls the conserved density of the bosons, $\rho = \Psi^\dagger \Psi$.
The coupling $g_0>0$ describes  a $U(2) = U(1)\times SU(2)$ invariant repulsive density-density contact interaction, as may be expected in a typical spinor BEC, while the $g_3$ parameter introduces anisotropy into the spin exchange interaction.
The $g_3$ coupling explicitly breaks the $SU(2)$ symmetry down to $ U(1)\otimes\mathbb{Z}_2$ comprised of rotations of the Bloch vector by any angle about the $z$ axis and reflections of the Bloch vector through the $xy$ mirror plane.
Note that stability requires that $g_3 > - g_0$.

Let us briefly comment that, while Lagrangian~\eqref{eqn:lagrangian} is a perfectly valid model, a more natural set-up may be realized by the more experimentally available spin-1 systems such as condensed $^7$Li, $^{23}$Na, or $^{87}$Rb.
All of these atoms are bosons which have a total hyperfine spin $F=1$ manifold~\cite{stamper-kurnSpinorBoseGases2013}.
In this case, the phase transition is between two phases which both respect the full $SU(2)$ spin-rotation symmetry\textemdash the ferromagnetic phase and polar (nematic) phase~\cite{hoSpinorBoseCondensates1998,ohmiBoseEinsteinCondensationInternal1998,barnettGeometricalApproachHydrodynamics2009}.
In this case, rather than being driven by anisotropy, the transition is driven by the overall sign of the spin-exchange interaction.
It turns out that the different ground-state phases have different types of Goldstone modes and therefore exhibit different analog spacetimes for the spin waves once condensate flow is introduced.
The relevant coupling constant is the spin-exchange coupling $c_2$, which is given in terms of the scattering lengths by
\[
c_2 = \frac{4\pi}{m} \frac{a_2 - a_0}{3}.
\]
For $^7$Li and $^{87}$Rb,$c_2< 0 $ while for $^{23}$Na $c_2 >0$~\cite{stamper-kurnSpinorBoseGases2013}.
Thus, all else equal we can realize both the polar (nematic) phase (which occurs for $c_2>0$) as well as the ferromagnetic phase ($c_2<0$) by using two different species of trapped atom.
All this is to say that, while Eq.~\eqref{eqn:lagrangian} is not as easily realized experimentally, there may be more experimentally feasible models which realize the same physics.
We now move on to the analysis of the technically simpler model proposed above.

The mean-field ground state of Eq.~\eqref{eqn:lagrangian} is identified as the homogeneous minimum of the energy density
\[
V =  \frac12 g_0 \left(\Psi^\dagger \Psi\right)^2 +\frac12g_3 \left(\Psi^\dagger \sigma_3 \Psi\right)^2 - \mu \Psi^\dagger \Psi.
\]
For $\mu <0$ the ground state is trivial and there is no condensate.
For $\mu >0$ there is Bose-Einstein condensation and the ground state is a BEC with a uniform condensate density which obeys the equation of state
\[
\rho = \Psi^\dagger \Psi =\begin{cases}
  \frac{\mu}{g_0}, & g_3>0, \\
  \frac{\mu}{g_0-|g_3|}, &  -g_0<g_3<0.
\end{cases}
\]
A non-zero condensate density always spontaneously break the overall $U(1)$ phase symmetry.
The corresponding Goldstone mode corresponds to the broken generator $i\sigma_0 = i\mathds{1}$ where $\mathds{1}$ is the $2\times 2$ identity matrix.

\begin{figure}
  \includegraphics{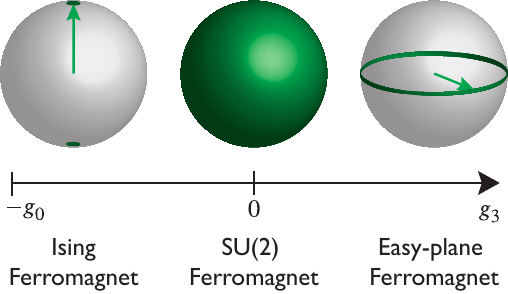}
  \caption{Illustration of the different ground-state Bloch-vector manifolds as the parameter $g_3$ is tuned.
  For $g_3 < 0$ the ground state manifold consists of the north and south poles and thus the system realizes an Ising ferromagnet, spontaneously breaking the $\mathbb{Z}_2$ symmetry while maintaining the $U(1)$ symmetry.
  For $g_3 =0 $ the full $SU(2)$ symmetry is realized and the ground-state manifold consists of the entire Bloch sphere.
  Thus, the system is a Heisenberg ferromagnet which spontaneously breaks the full $SU(2)$ down to $U(1) \subset SU(2)$.
  Finally, for $g_3 > 0$ the ground state manifold consists of the equatorial plane, rendering the system an XY (easy-plane) ferromagnet.
  Thus, the initial symmetry is $U(1)$  which is spontaneously broken to the trivial group.
  }
  \label{fig:phasediagram}
\end{figure}

\begin{table}
  \begin{ruledtabular}
  \begin{tabular}{lcc}
    Phase & Sound waves & Spin waves \\
    \hline
    Ising Ferromagnet & $\omega \sim k$ & Gapped \\
    SU(2) Ferromagnet & $\omega \sim k$ & $\omega \sim k^2$ \\
    Easy-plane Ferromagnet & $\omega \sim k$ & $\omega \sim k$
  \end{tabular}
  \end{ruledtabular}
  \caption{Goldstone modes associated to each phase shown in Fig.~\ref{fig:phasediagram}.
  All phases have a Type-I Goldstone mode associated to the spontaneous breaking of the global $U(1)$ phase, corresponding to the conventional sound mode.
  Additionally, there may also be Goldstone modes associated with spontaneous breaking of spin symmetries, leading to spin waves.
  In the Ising phase, the broken symmetry is discrete and there are no Goldstone modes.
  In the SU(2) invariant Heisenberg phase there is a Type-II Goldstone mode describing transverse fluctuations of the magnetization, while in the XY easy-plane phase there is a Type-I Goldstone describing equatorial fluctuations of the magnetization.}
  \label{tab:mode-properties}
\end{table}

Depending on the value of $g_3$, additional symmetries may be broken, resulting in the phase diagram illustrated in Fig.~\ref{fig:phasediagram}.
We write the condensed $\Psi$ in the density-phase-spinor representation as
\begin{equation}
    \label{eqn:spinor-form}
    \Psi = \sqrt{\rho}e^{i\Theta} \chi, \quad \chi^\dagger \chi = 1
\end{equation}
where $\chi$ yields the local magnetization density.
It may be parameterized in terms of one complex parameter $\zeta$ via
\begin{equation}
\chi = \frac1{\sqrt{2(1+|\zeta|^2)}}\begin{pmatrix}
1 + \zeta \\
1 - \zeta
\end{pmatrix}, \quad \zeta \in \mathbb C.
\end{equation}
Alternatively, it may be represented in the more canonical Euler angle representation as
\[
\chi = \begin{pmatrix}
\cos\tfrac{\theta}{2} \\
\sin\tfrac{\theta}{2}e^{i\varphi}
\end{pmatrix}, \quad \varphi \in [0,2\pi)\quad \theta \in [0,\pi).
\]
We use both of these representations throughout.
In terms of $\zeta$ and $\theta,\varphi$ the anisotropic interaction is
\[
V = \frac12 g_3 \rho^2 \frac{(\zeta + \zeta^*)^2}{(1 + |\zeta|^2)^2} = \frac12 g_3 \rho^2 \cos^2 \theta.
\]
We now proceed to study the mean-field phase diagram of the ground state.

{\it Ising phase.}\textemdash We begin by considering the case of $g_3 <0$, i.e. the ``Ising ferromagnet" phase.
The interaction has a $U(1)\times \mathbb{Z}_2$ symmetry generated by $\tfrac{i}{2} \sigma_3$ composed with inversion of the $z$ component of the magnetization.
In this case it is energetically favorable for the Bloch vector to align with the $z$ axis.
This breaks the $\mathbb{Z}_2$ symmetry and preserves $U(1)$ so the ground state manifold is the symmetric space $U(1)\times \mathbb{Z}_2 / U(1) \sim \mathbb{Z}_2$.
This is depicted in the left-most panel of Fig.~\ref{fig:phasediagram}, which shows the ground-state manifold for the spinor $\chi$ for various couplings.
The Goldstone modes associated with the broken-symmetry ground-state, along with their dispersions are shown in Table~\ref{tab:mode-properties}.
As the ground-state manifold is discrete there is no additional Goldstone mode in this phase and we no longer consider this portion of the phase diagram in this work.

{\it Heisenberg phase.}\textemdash When $g_3 =0$ the interaction term is isotropic and the model has the full $SU(2)$ invariance.
The ground state then spontaneously break the $SU(2)$ symmetry down to $U(1)$ so that the ground state manifold is the symmetric space $SU(2)/U(1)\sim S^2$\textemdash the full Bloch sphere.
This is illustrated in the middle panel of Fig.~\ref{fig:phasediagram}.
Without loss of generality, we take the ground state magnetization to point along the positive $x$ direction.
Thus, $\zeta = 0$ and $\chi = \frac{1}{\sqrt{2}}(1,1)^T$.
Then the unbroken generators are $\{\tfrac{i}{2}( \sigma_1 - \mathds{1} )\}$ and the broken generators are $\{\tfrac{i}{2}(\sigma_1 + \mathds{1}), \tfrac{i}{2}\sigma_2 , \tfrac{i}{2}\sigma_3, \}$.
Using the formalism from Sec.~\ref{sec:spacetimes}, we find that the $P$ matrix appearing in the Goldstone mode Lagrangian is
\begin{equation}
  P^t = \rho \begin{pmatrix}
    0 & 0 & 0 \\
    0 & 0 & \tfrac14 \\
    0 & -\tfrac14 & 0
\end{pmatrix},
\end{equation}
where the columns refer, in order, to the generators $\{\tfrac12 i \sigma_0 + \tfrac12 i \sigma_1, \tfrac12 i \sigma_2, \tfrac12 i\sigma_3\}$.
In this case, we have one Type-II Goldstone mode associated with the two generators $\{\tfrac12 i \sigma_2, \tfrac12 i\sigma_3\}$ which exhibits a quadratic dispersion relation and hence realize the Newton-Cartan geometry in the presence of inhomogeneous condensate flow.
This is summarized in Table~\ref{tab:mode-properties}.

{\it XY phase.}\textemdash We now move on to the case where $g_3>0$.
In this case there is an energy penalty associated with a non-zero $z$ component of the magnetization and thus the ground state lies in the manifold defined by $\cos \theta = 0 \Rightarrow \theta = \pi/2$.
Thus, the ground state breaks the $U(1)$ symmetry but remains invariant under reflections through the $z=0$ plane.
As such, the ground state resides in the symmetric space $U(1)\times \mathbb{Z}_2 / \mathbb{Z}_2 = U(1) \sim S^1$, as depicted in the right panel of Fig.~\ref{fig:phasediagram}.
Without loss of generality we again take the Bloch vector to lie along the $+x$ direction.
Thus, only two generators remain unbroken in the Lagrangian $\{i \mathds{1}, \tfrac12 i\sigma_3\}$ and the mean-field breaks both of them.
We again refer to Eq.~\eqref{eq:typeiandtypeiimatrix} to obtain
\begin{equation}
  P^t_{mn} = 0.
\end{equation}
Thus, there are no Type-II Goldstone modes in this system, but instead two Type-I modes which are linearly dispersing and therefore exhibit an analog Einstein-Hilbert spacetime, summarized in Table~\ref{tab:mode-properties}.

\subsection{Bogoliubov-de Gennes Analysis}\label{subsec:bdg}

We now proceed to examine the fluctuations about the mean-field by obtaining and diagonalizing the Bogoliubov-de Gennes equations of motion.
To see how the analog spacetime emerges we consider a mean-field condensate $\psi_0$ which is inhomogeneous, but has a constant magnetization density.
Taking the spin to point in the $+x$ direction, we obtain
\begin{equation}
    \label{eqn:gs-mf}
    \psi_0 = \sqrt{\rho(x)}e^{i\Theta(x)}\chi_0 = \sqrt{\rho(x)}e^{i\Theta(x)} \begin{pmatrix}
    \frac{1}{\sqrt{2}}\\
    \frac{1}{\sqrt{2}}\\
    \end{pmatrix} .
\end{equation}
In this case, the mean-field describes a flowing condensate with superfluid density $\rho(x) = \psi_0^\dagger(x) \psi_0(x)$ and superfluid velocity $\mathbf{v}_s = \frac{1}{m} \nabla \Theta(x)$.
Fluctuations about this mean-field can be fully parameterized in terms of the two complex fields $\phi$ and $\zeta$ as
\begin{equation}
\label{eqn:fluc-form}
\delta \Psi = \left(\phi \sigma_0 +  i\zeta \sigma_2\right)\psi_0.
\end{equation}
To quadratic order, the Lagrangian from Eq.~\eqref{eqn:lagrangian} decouples into two quadratic BdG Lagrangians
\begin{align}
  \mathcal L_\phi & = \rho \left[ \tfrac i2 (\phi^* D_t \phi - \phi D_t \phi^*) - \tfrac{|\nabla \phi|^2}{2m} + \tfrac12 g_0 \rho (\phi + \phi^*)^2 \right],\nonumber \\
  \mathcal L_\zeta & = \rho \left[ \tfrac i2 (\zeta^* D_t \zeta - \zeta D_t \zeta^*) - \tfrac{|\nabla \zeta|^2}{2m} + \tfrac12 g_3 \rho (\zeta + \zeta^*)^2 \right],\label{eqn:Bdg-Lagrangians}
\end{align}
with $D_t = \partial_t + \mathbf{v}_s \cdot \nabla$ the material derivative in the frame co-moving with the superfluid flow.
These two Lagrangians are specific examples of the more general Eq.~\eqref{eq:full-fluc-Lagrangian}.
In particular, for $g_3 > 0$ at long wavelengths we can apply the analysis of Sec.~\ref{sec:relativistic} to obtain the relativistic analog spacetime.
If on the other hand, $g_3 = 0$, then at long wavelengths we can apply the analysis of Sec.~\ref{sec:nonrelativistic} to obtain the nonrelativistic Newton-Cartan analog spacetime.
Nevertheless, it is instructive to instead follow Ref.~\cite{curtisEvanescentModesSteplike2019,recatiBogoliubovTheoryAcoustic2009}, and directly employ the BdG equations when determining the consequences of the changing spacetime structure.
This is because the BdG equations provide us with a single unified description with which we may capture both phases, as well as the transition between them.

The BdG equations are obtained as the Euler-Lagrange equations of Lagrangians $\mathcal{L}_{\phi}, \mathcal{L}_{\zeta}$ and are most transparently expressed in terms of the Nambu spinors
\begin{equation}
    \label{eqn:Nambu}
    \Phi_0 = \left(\begin{array}{c}
    \phi \\
    \phi^*\\
    \end{array}\right), \ \Phi_3 = \left(\begin{array}{c}
    \zeta\\
    \zeta^*\\
    \end{array}\right)
\end{equation}
for condensate and spin wave fluctuations, respectively.
We then find the BdG equations $\hat{K}_0 \Phi_0 = 0$, and $\hat{K}_3 \Phi_3 = 0$, with the BdG differential operators
\begin{equation}
    \label{eqn:BdG-EOM}
    \begin{aligned}
    &   \hat{K}_{0} = \tau_3 \left( i \partial_t +i\mathbf{v}_s\cdot\nabla\right) + \frac{1}{2m\rho}\nabla\cdot\rho\nabla\tau_0 - g_0\rho\left(\tau_0 + \tau_1\right) \\
    &   \hat{K}_{3} = \tau_3 \left( i \partial_t +i\mathbf{v}_s\cdot\nabla\right) + \frac{1}{2m\rho}\nabla\cdot\rho\nabla\tau_0 - g_3\rho\left(\tau_0 + \tau_1\right), \\
    \end{aligned}
\end{equation}
written in terms of the Nambu particle-hole Pauli matrices $\tau_a$.
Let us emphasize that the only difference between $\hat{K}_0$ and $\hat{K}_3$ is the coupling constant appearing in front of the $\tau_0+\tau_1$ term.
For sound waves it is $g_0$, while for the spin waves it is $g_3$.
Thus, both Goldstone modes end up coupling to the same background condensate density and velocity, albeit with different speeds of sound.
Sound waves end up propagating with the local group velocity
\[
c_0(x) = \sqrt{\frac{g_0 \rho(x)}{m}}
\]
while the spin waves have the local group velocity
\[
c_3(x) = \sqrt{\frac{g_3\rho(x)}{m}}.
\]
Thus, we see that the coupling $g_3$ allows us to independently tune the two speeds of sound relative to each other.

For generic values of $g_3>0$ and arbitrary condensate flows we cannot find quantum numbers with which we can diagonalize $\hat{K}_3$.
However, at the $SU(2)$ symmetric point $g_3 =0$ we observe that the BdG kernel for spin waves obeys
\[
\hat{K}_{3} = \tau_3 \left( i \partial_t +i\mathbf{v}\cdot\nabla\right) + \frac{1}{2m\rho}\nabla\cdot\rho\nabla\tau_0 \Rightarrow \left[ \tau_3, \hat{K}_3 \right] = 0.
\]
Since $\tau_3$ now commutes with the kernel, the two components of the BdG spinor decouple and each independently obeys a Galilean-invariant dispersion relation.
This also results in an additional $U(1)$ symmetry generated by $\tau_3$ which imposes a selection rule for the allowed Bogoliubov transformations.
In particular, there is no matrix element which scatters a ``particle-like" Bogoliubov quasiparticle into a ``hole-like" particle.
this process is the one responsible for Hawking radiation and as such we find, counter-intuitively, that it is impossible to generate Hawking radiation in the Newton-Cartan spacetime despite the fact that all flow velocities $\mathbf{v}_s$ are now supersonic.
This is explicitly demonstrated for the case of a step-like horizon, which we analyze in the following section.

\section{Step-Like Horizon}\label{sec:step}

In order to get a more quantitative understanding of how the changing spacetimes affect observable physics, we imagine a specific flow profile and use the BdG equations to solve for the spin-wave scattering matrix.
We imagine a quasi-one-dimensional stationary condensate flow with a superfluid density and velocity which obeys $\partial_t \rho = \partial_t v_s = 0$.
The continuity equation for the condensate then implies
\begin{equation}
    \label{eqn:continuity}
    \partial_x(\rho v_s) = 0 \Rightarrow \rho(x) v_s(x) = \textrm{const}.
\end{equation}
The local speed of sound for the spin-waves (henceforth simply written as $c$) is therefore $c(x) = \sqrt{g_3 \rho(x)/m}$.

To further simplify calculations, we consider the case of a step-like profile for $\rho(x),v(x)$ of the form
\begin{equation}
    \label{eqn:step-flow}
\begin{aligned}
   & \rho(x) = \left\{\begin{aligned}
    &\rho_l   &   x < 0 \\
    &\rho_r   &   x \geq 0 \\
\end{aligned}\right.  \\
& v(x) = \left\{\begin{aligned}
    & - |v_l|   &   x < 0 \\
    & - |v_r|   &   x \geq 0 .\\
\end{aligned}\right. \\
\end{aligned}
\end{equation}
Note that continuity requires $v_l \rho_l = v_r \rho_r \Leftrightarrow v_lc_l^2 = v_r c_r^2$.
In this work we adopt the convention that $v$ is negative, so that the condensate flows from the right to the left.
With this set-up, we can employ the BdG techniques usually used for phonon modes to these spin waves~\cite{recatiBogoliubovTheoryAcoustic2009,curtisEvanescentModesSteplike2019}.

This step-like potential has the advantage that away from the jump, momentum eigenstates solve the BdG equations, and the scattering matrix reduces to a simple plane-wave matching condition at the boundary.
The details of this procedure may be found, e.g. in Appendix~\ref{app:BogoliubovHawking}.
Here we simply discuss the results of the calculation.
We start by considering $g_3 >0$ to be large and then decrease down to zero.
As we do so, while keeping the flow profile fixed, we pass through three regimes.

The first regime occurs for large $g_3$ so that $c_l>|v_l|$ and $c_r > |v_r|$.
Thus, there is no sonic horizon and no Hawking radiation.

Eventually as we continue decreasing $g_3$ we enter the regime where $|v_r| < c_r$ but $c_l<|v_l|$.
This exhibits a sonic horizon at $x=0$ and is thus accompanied by Hawking radiation.

Finally, we reach the regime where $|v_l| > c_l $ and $|v_r|> c_r$.
This is a novel regime wherein both the interior and exterior of the jump are supersonic.
However, due to the non-linear Bogoliubov dispersion, there are still some short-wavelength modes for which one or both sides of the flow are not supersonic (this is due to the convex dependence of the group-velocity on momentum).
Thus there is still Hawking radiation, however we find that as we decrease $g_3$ further, the total ``flux" of modes which are emitted decreases until we recover the result that at $g_3=0$ there is no radiation at all.

To see this, we define the ``total number of Hawking modes" at a given frequency to be $N(\omega)$ (see Eqs.~\eqref{eq:Nomega1} and \eqref{eq:Nomega2}).
This is obtained by calculating the ``Hawking" element of the scattering matrix for the BdG equations.
From $N(\omega)$ we can then define the total ``luminosity'' \cite{corleyHawkingSpectrumHigh1996} leaving the horizon by
\begin{equation}
    L_{\mathrm{H}} = \int_0^\infty d\omega \ \frac{\omega}{2\pi} N(\omega).
\end{equation}
Note that in the conventional black hole case, $N(\omega)$ is the number of photons at frequency $\omega$ seen at asymptotic infinity and thus this is simply the number flux per unit frequency of the radiation.

\begin{figure}
    \centering
    \includegraphics{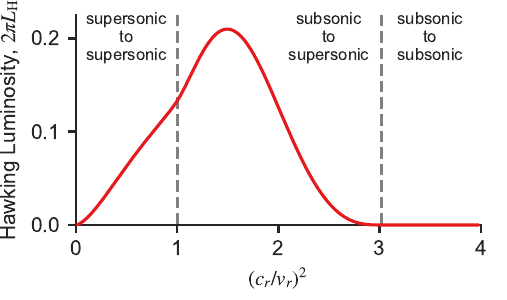}
    \caption{The total luminosity due to the Hawking radiation for a fixed density profile $\rho(x)$ and velocity profile $v(x)$.
    We see that there is no Hawking radiation when $c_r$ is sufficiently large so that  $c_l>v_l$ (recall these are constrained by the continuity equation).
    When $c_l<v_l$ but $c_r> v_r$ we get a region of subsonic flow that flows into a supersonic region and we begin seeing traditional Hawking radiation.
    As we further tune $g_3$, $c_r$ drops below $v_r$ and both regions become supersonic at low frequencies.
    Evidently, there is still a channel for Hawking radiation emission as seen by the non-zero integrated flux.
    However, as $c_r$ drops to zero this channel closes, vanishing precisely at the quantum phase transition into the Newton-Cartan geometry ($c_r=0=g_3$).
    In this plot, {$v_l = 1.3$, $v_r=0.9$}, $m=10$, and $\rho(x)v(x)=1$.}
    \label{fig:totalflux}
\end{figure}

The upshot is given by Fig.~\ref{fig:totalflux} which plots $L_\mathrm{H}$ as a function of $(c_r/v_r)^2 = g_3 \rho_r /m v_r^2$.
Thus, for fixed flow density and velocity, this is essentially plotting as a function of the control parameter $g_3$.
We see the three distinct regions and importantly at $g_3=0$ we see the Hawking effect vanish.

To understand this effect, we consider the dispersion relation of the waves away from the horizon, for which momentum is a good quantum number.
In the right and left half-spaces we have the relations
\begin{equation}
    (\omega - v_{\alpha} k)^2  =  c_{\alpha}^2 k^2 + \frac{k^4}{4 m^2},
\end{equation}
where $\alpha = l,r$ for the left and right regions respectively.
This relates the lab-frame frequency of a wave $\omega$ to the lab-frame momentum $k$.
This dispersion relation is plotted in Figs.~\ref{fig:SuperToSubHawking} and \ref{fig:SuperToSuperHawking}.
Due to the presence of a discontinuity at $x = 0$ modes with different momenta mix and only $\omega$ can be fixed globally.
Thus, the dispersion relation is to be solved by finding the allowed momenta at each fixed lab-frame frequency.
This amounts to finding the roots of a quartic polynomial with real coefficients, and as such there are always four solutions (which are either real or complex conjugate pairs).
The real momenta represent propagating modes while we later find that the complex roots describe evanescent modes localized around the horizon.

\subsection{Subsonic-Supersonic Jump}

\begin{figure}
  \includegraphics{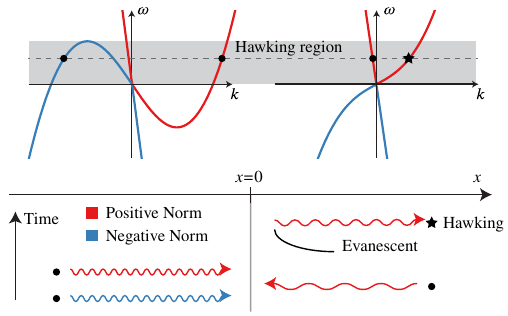}
  \caption{The Hawking effect for $g_3$ such that $c_r> v_r$ and $c_l < v_l$ (sub-sonic to super-sonic).
  In this situation, one side (left) flows faster than the speed of some excitations, and the other side (right) flows slower than the speed of any excitation.
  The dashed line represents the constant lab frame energy $\omega$.
  The mode that carries away energy from the horizon is the ``Hawking mode,'' shown by the star marker.
  Tracing this mode back in time (bottom of figure), we find that it comes from a scattering process that includes positive (red) and negative (blue) norm states.
  It is the negative norm state to the left of the horizon that is responsible for particle creation in the Hawking channel.
  Notice that for frequencies larger than those in the labeled ``Hawking region,'' there is no Hawking effect due to lack of negative energy modes to have scattered from at earlier times. }
  \label{fig:SuperToSubHawking}
\end{figure}

First, we consider the case of a jump between a subsonic and supersonic flow, depicted graphically in Fig.~\ref{fig:SuperToSubHawking}.
In this case, we recover the well-known result that there is Hawking radiation emitted.
The dispersion relation in each half-plane is plotted and intercepts with a constant $\omega>0$ are found.
These intercepts yield the momenta of the propagating modes in each region for the given frequency.
Each curve is depicted with a color indicating the sign of the group velocity in the {\bf co-moving} frame, which is what is used to distinguish between ``particle-like" (red) and ``hole-like" (blue), in accordance with the BdG norm (see Appendix~\ref{app:BogoliubovHawking} and in particular Eq.~\eqref{eq:inner-product} for definition).
We see that the outgoing Hawking mode (combined with an evanescent piece at the horizon) is connected to three incoming waves, one of which is a negative norm state originating from the interior of the horizon.
This particle-hole conversion processes is the origin of the Hawking effect, as this induces a Bogoliubov transformation which connects the vacuum of the asymptotic past to a one-particle state in the asymptotic future (and vice-versa).

We see that due to the convex non-linear Bogoliubov dispersion relation, there is a maximum frequency of the emitted Hawking radiation obtained by finding the local maximum of the negative norm dispersion relation.
Above this frequency, the flow is no longer supersonic since the group velocity of modes depends non-trivially on the frequency.

\subsection{Supersonic-Supersonic Jump}
\begin{figure}
  \includegraphics{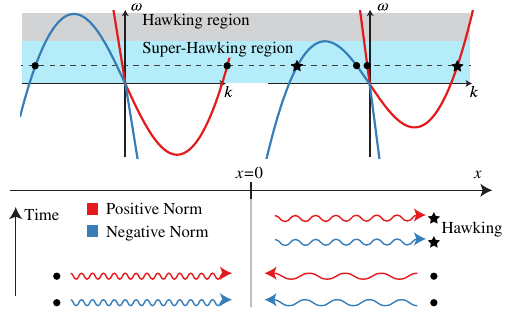}
  \caption{The Hawking effect for $g_3$ such that $c_r< v_r$ and $c_l < v_l$ (super-sonic to super-sonic). With both regions flowing faster than the speed of excitations (relative to the horizon), we still have a Hawking region, but now we also have a ``Super-Hawking'' region where the positive and negative normalization modes from both regions can scatter between one another.}
  \label{fig:SuperToSuperHawking}
\end{figure}

\begin{figure}
  \includegraphics[width=\columnwidth]{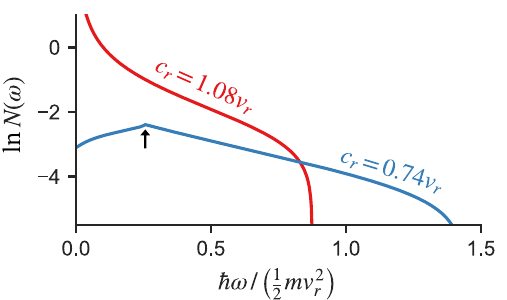}
  \caption{Hawking flux $N(\omega)$ as a function of frequency for the subsonic-to-supersonic case (red) and the supersonic-to-supersonic case (blue). As we approach the Heisenberg symmetric point $g_3=0$, we find the Hawking flux disappears both in its overall magnitude and singular behavior.
  The black arrow indicates the onset of the ``super-Hawking region'' responsible for the absence of the singular distribution.
  }
  \label{fig:HawkingFlux}
\end{figure}

As we decrease $g_3$ beyond a critical value the system enters a parameter regime where both sides of the jump are supersonic flows.
In this case, the dispersion relation still exhibits a Hawking-like region, as we see in Fig.~\ref{fig:SuperToSuperHawking}.
However, we also see a new region emerge at low energies (labeled ``super-Hawking'' in the figure) in which now both a positive and negative norm mode can be scattered into.
This opens a new channel in the scattering matrix which leads to a reduction in the amplitude for scattering into the Hawking channel, as per generalized unitarity constraints.
This is seen in Fig.~\ref{fig:HawkingFlux}, which compares $N(\omega)$ for the case of a subsonic-supersonic (red) and supersonic-supersonic jump (blue).
Both curves are qualitatively similar at high frequencies, corresponding to the ``Hawking" region of frequencies in Figure~\ref{fig:SuperToSuperHawking}.
On the other hand, we see that at low $\omega$, when we have subsonic-to-supersonic flow, $N(\omega)$ diverges in the universal thermal manner, while in the supersonic-to-supersonic regime, there is a noticeable change in behavior between the Hawking and super-Hawking regimes, cutting off this low $\omega$ divergence.

There are two effects occurring which are responsible for decreasing the Hawking luminosity $L_\mathrm{H}$.
First, in the Hawking region the incoming negative norm states now begin to more strongly backscatter into their corresponding negative norm state, occupying the evanescent mode on the right side of the horizon.
Second, in this super-Hawking region, the appearance of an outgoing negative-norm mode provides an opportunity for the ingoing negative norm channel to avoid scattering into the positive norm channel.
We indeed find that the two channels begin to decouple from each other, diminishing the amount of Hawking radiation that can be produced.

\subsection{Absence of Hawking Radiation for Type-II modes\label{subsec:noHawking}}
This takes us directly into the point where $g_3 = 0$, which exhibits the new Newton-Cartan spacetime geometry.
One might expect that there should be something akin to a Hawking effect since some modes ``see'' a horizon for any difference in $|v_l|$ and $|v_r|$.
However, this horizon does not translate into a Hawking effect.
As explained earlier, at this point the BdG kernel $\hat{K}_3$ commutes with $\tau_3$.
In terms of the BdG Lagrangian of Eq.~\eqref{eqn:Bdg-Lagrangians}, we find that there is now a new global $U(1)$ symmetry $\zeta \rightarrow e^{i\vartheta} \zeta$.
We can see explicitly from the BdG analysis that this conserved charge density is given by
\[
Q_{\mathrm{BdG}} = \int d^3 x \, \rho |\zeta|^2.
\]
On the other hand, by applying Noether's theorem directly on the general Newton-Cartan action of Eq.~\eqref{eq:NC-lagrangian-II}, in the limit where $n_0$ is the only nonzero component of $n_\mu$ and the Lagrangian is independent of the $x^0$, we find
\begin{equation}
    Q_{\mathrm{BdG}} = \int d^3 r \sqrt{h}|\psi|^2 . \label{eq:NCconservedQ}
\end{equation}
If we identify $\psi = \zeta$ and use the results of Eq.~\eqref{eqn:geometry-result} we find that these two indeed match each other.
In particular, Eq.~\eqref{eq:NCconservedQ} describes a conserved charge for the field $\psi$ on a curved manifold given by $h^{\mu\nu}$.

\begin{figure}
  \includegraphics[width=\columnwidth]{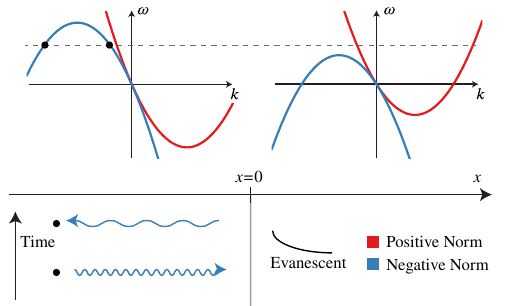}
  \caption{For $g_3 = 0$ in the Newton-Cartan geometry there is an excitation number conservation that protects negative norm states from scattering into positive norm states and as a result, if we scatter a negative norm state in what used to be the ``Hawking region,'' we find it fully back scatters into a negative norm state and leaks past the horizon only with an evanescent tail characteristic to a ``classically forbidden'' region.}
  \label{fig:ksqr}
\end{figure}

Since, unlike the charge in Eq.~\eqref{eq:inner-product}, this density is positive definite it can be genuinely interpreted as the number of BdG quasiparticles.
This symmetry then imposes a selection rule on the scattering matrix which prohibits the scattering processes responsible for the Hawking process, which leads to a creation of BdG quasiparticles.
This is evident if we see that when $g_3 = 0$,
\begin{equation}
 \left[ i\left(\partial_t + \mathbf{v}\cdot\nabla\right)+ \frac{1}{2m\rho}\nabla\cdot\rho\nabla \right]\zeta = 0,
\end{equation}
and hence $\zeta$ and $\zeta^*$ do not mix.
Indeed, as Fig.~\ref{fig:ksqr} illustrates, though Hawking radiation is permissible by conservation of energy and momentum, as seen by the dispersion relation in Fig.~\ref{fig:ksqr}, there is no permissible matrix element for any scattering process which mixes positive and negative norm modes.
Thus, at low frequencies (below the cutoff frequency on the right), negative norm modes may be transmitted across the horizon but only as outgoing negative norm modes.
This is analogous to the ``super-Hawking" regime earlier, but since there is no conversion between positive and negative norm modes, there is no Hawking radiation effect.

Above the cutoff frequency on the right (in what we refer to as the ``regular Hawking regime"), all negative norm modes incident from the interior of the horizon must be reflected back.
Even in this case, there is still a finite penetration of the negative norm state across the event horizon in the form of an evanescent mode which is decaying away from the horizon, as originally predicted in Ref.~\cite{curtisEvanescentModesSteplike2019}.
In fact, this evanescent tail is also present when $g_3 > 0$, but now it is not accompanied by any other outgoing mode.
Again, let us emphasize that this evanescent mode is associated with a negative norm mode and therefore does not couple to positive norm modes.
Thus, it cannot be spontaneously excited from the ingoing vacuum.
Ultimately, as the negative norm mode must be reflected, all the amplitude which initially went into the outgoing positive norm states when $g_3 >0$ is now transferred into the reflected negative norm state and the evanescent tail.

% m=10
% ζ=1.0
% vL=1.3
% cL=sqrt(ζ/vL)
% vR=0.9
% cR=sqrt(ζ/vR)

\section{Transport in Newton-Cartan Geometry\label{sec:transport}}

In this section we take up the issue of energy transport in systems exhibiting Newton-Cartan geometry.
Building on Luttinger's work on computing heat transport via coupling to a gravitational field~\cite{luttingerTheoryThermalTransport1964}, there has been a well-established method of coupling systems to Newton-Cartan geometry in order to extract their heat transport properties~\cite{gromovThermalHallEffect2015,son2013newtoncartan,bradlynLowenergyEffectiveTheory2015,geracieSpacetimeSymmetriesQuantum2015}.
With these methods, we can begin with the results in Sec.~\ref{sec:nonrelativistic} and find the stress tensor $T^{\mu\nu}$, energy current $\epsilon^\mu$, and momentum density $p_\mu$.
However, as we have mentioned previously, we can also reformulate the relativistic Lagrangian in Sec.~\ref{sec:relativistic} in terms of a Newton-Cartan geometry with an additional external field.
Therefore, in the bulk of this section, we make that precise and use the energy transport machinery to relate the relativistic stress-energy tensor of Type-I modes to its non-relativistic counterparts.

We begin by noting that the variations in the geometry are not independent as they must satisfy the constraints imposed by Newton-Cartan geometry that $n_\mu v^\mu = 1$ and $n_\mu h^{\mu\nu} = 0$.
Parameterizing the variations so as to respect these constraints is done by introducing the perturbations $\delta n_\mu$, $\delta u^\mu$ and $\delta \eta^{\mu\nu}$ such that
\begin{equation}
    \begin{split}
        \delta v^\mu & = - v^\mu v^\lambda \delta n_\lambda + \delta u^\mu, \\
        \delta h^{\mu \nu} & = - (v^\mu h^{\nu \lambda} + v^\nu h^{\mu \lambda}) \delta n_\lambda - \delta \eta^{\mu\nu},
     \end{split}
\end{equation}
where $n_\mu \delta u^\mu = 0$, and $n_\mu \delta \eta^{\mu \nu} = 0$ so that $\delta u^\mu$ and $\delta \eta^{\mu\nu}$ are orthogonal to the clock one-form $n_\mu$.

To find the full Lagrangian it is useful to formally define a non-degenerate metric in the full spacetime by
\begin{equation}
    g^{\mu\nu} \equiv v^\mu v^\nu + h^{\mu\nu}.
\end{equation}
Note that unlike relativistic metrics, this Newton-Cartan has no invariant distinction between space-like and time-like separations (simultaneity is a global concept imposed by $n_\mu$).
As $g^{\mu\nu}$ is non-degenerate, we may proceed to take the inverse which is defined by
\begin{equation}
    g_{\mu \alpha} g^{\alpha \nu} = \delta_\mu^\nu,
\end{equation}
where $\delta_\mu^\nu$ is the usual Kronecker delta.
This also serves to define the inverse of the degenerate metric $h^{\mu\nu}$ by
\begin{equation}
    g_{\mu\nu} \equiv n_\mu n_\nu + h_{\mu\nu}.
\end{equation}
Note that the constraints on the geometry then imply $h_{\mu\nu}$ obeys
\begin{equation}
    h^{\mu\sigma}h_{\sigma\nu} = \delta^{\mu}_\nu - v^\mu n_\nu.
\end{equation}
The right hand side essentially acts to project onto the manifold upon which $h^{\mu\nu}$ is not degenerate.
These are the ``spatial" three-surfaces which are in some sense ``iso-temporal."

Introducing $g$ is helpful in particular because we then find that if take the determinant $g= \det(g_{\mu\nu})$, we find that $\sqrt{g} = n_0 \sqrt{h}$ \footnote{
This is derived more directly using $g^{-1}$ defined by $g^{\mu\nu}$. If one locally takes $n_\mu = (n_0, \mathbf 0)$, then $g^{00} = (v^0)^2 \equiv A^{00}$, $g^{0i} = g^{i0} = v^0 v^i \equiv B^{0i}$ and $g^{ij} = v^i v^j + h^{ij} \equiv D^{ij}$. One can take the Schur complement of this inverse metric $g^{-1}/A$ to compute the determinant; then $1/g = \det(g^{-1}) = \det(A) \det(D - B^T A^{-1} B) = 1/(n_0^2 h)$.
}.
This is exactly the volume measure of the Lagrangian Eq.~\eqref{eq:NC-lagrangian-II}.
This assists in taking the variation
\begin{equation}
    \delta[\sqrt{g}] = \sqrt{g}[v^\mu \delta n_\mu + \tfrac12 h_{\mu\nu}\delta\eta^{\mu\nu}].
\end{equation}
We can then use the variations to find the stress tensor $T_{\mu\nu}$, energy current $\epsilon^{\mu}$, and momentum density $p_\mu$ via \cite{geracieSpacetimeSymmetriesQuantum2015}
\begin{equation}
    \delta S = \int d^{d+1} x \sqrt{g} \left(\tfrac12 T^{\mu\nu} \delta \eta_{\mu \nu} - \epsilon^\mu \delta n_\mu - p_\mu \delta u^\mu \right).
\end{equation}
Due to the constraints on $\delta u^\mu$ and $\delta \eta_{\mu\nu}$, these values of $p_\mu$ and $T^{\mu\nu}$ are not unique.
In fact, we can make any substitution $p_\mu\rightarrow p_\mu + a n_\mu$ or $T^{\mu\nu}\rightarrow T^{\mu\nu} + b^\mu v^\nu + b^\nu v^\mu$.
We impose uniqueness by requiring $p_\mu v^\mu = 0$ and $T^{\mu\nu} n_\nu = 0$.
Lastly, one can derive continuity equations for these quantities by considering how these objects change under a diffeomorphism (see Ref.~\cite{geracieSpacetimeSymmetriesQuantum2015}).

We now compute these quantities for both the Type-I and Type-II modes.
It is worth noting that these models describe the free propagation of Goldstone modes and thus are in a sense ``non-interacting."
By this, we mean there are no additional terms due to interactions~\cite{liao2019drag}.
For Type-II modes, the resulting transport quantities are known~\cite{son2013newtoncartan,bradlynLowenergyEffectiveTheory2015,gromovThermalHallEffect2015,jensenCouplingGalileaninvariantField2018,jensenAspectsHotGalilean2015}.
We briefly recapitulate this calculation here.

\subsection{Energy transport for Type-II modes}

We proceed to vary the Newton-Cartan geometry in action Eq.~\eqref{eq:NC-lagrangian-II}.
This straightforwardly yields the momentum density as
\begin{equation}
    p_\mu = -\tfrac{i}{2} \left[ \bar{\psi} (\partial_\mu - n_\mu v^\alpha \partial_\alpha) \psi -  \psi (\partial_\mu - n_\mu v^\alpha \partial_\alpha) \bar{\psi} \right].
\end{equation}
The limit works out as expected: if we let $n_\mu = (1,\mathbf{0})$ and $v^\mu = (1,\mathbf{0})^T$, only the spatial components survive and we obtain the momentum current for a non-relativistic theory with conserved density $|\psi|^2$.
Next, we compute the stress tensor, which describes the momentum flux.
We find
\begin{multline}
T^{\mu\nu} =  -\tfrac{i}{4} v^\alpha\left[ \bar{\psi} \partial_\alpha  \psi -  \psi \partial_\alpha \bar{\psi} \right] h^{\mu \nu} \\ + \tfrac1{4m} \partial_\alpha \bar\psi \partial_\beta \psi (h^{\alpha\mu} h^{\beta \nu} + h^{\alpha\nu} h^{\beta \mu} - h^{\mu\nu} h^{\alpha\beta})
\end{multline}
and the energy current as
\begin{equation}
\epsilon^\mu = -\tfrac{1}{2m}(\partial_\alpha \bar{\psi})(\partial_\beta \psi) \left[ v^\alpha h^{\beta \mu} + v^{\beta}h^{\alpha\mu} -v^\mu h^{\alpha\beta} \right].
\end{equation}
Both have sensible flat-space limits as well.

\subsection{Energy transport for Type-I modes}

For Type-I modes, an analog relativistic theory emerges from a nonrelativistic theory, and in both the cases, we can compute energy densities, momentum densities, and the stress-tensor.
The objective of this section is to compute how the quantities in the analog relativistic system are related to their nonrelativistic counterparts, motivated by the spacetime relations derived in Sec.~\ref{sec:spacetimes}.

We have shown the Type-I modes can be thought of as residing in a relativistic analog spacetime, equipped with an analog metric tensor $\mathcal{G}_{\mu\nu}$.
If we vary with respect to this tensor, we obtain a Lorentz-invariant stress-energy-momentum tensor, $\mathcal{T}^{\mu\nu}$.
Note Lorentz invariance constrains this to be symmetric, relating the energy current and momentum densities to each other.

On the other hand, we have shown that one can obtain the Type-I modes by gapping out one of the generators of a Type-II mode.
Thus, we can also consider varying the Newton-Cartan geometry that the Type-II mode resides in before including a mass gap.
This yields for us the Newton-Cartan stress tensor, momentum density, and energy current and provide for us a general relationship between the relativistic energy-momentum tensor and the non-relativistic counterparts.

First, we return to Eq.~\eqref{eqn:type-i-fluc} and rewrite the Lagrangian in terms of the Newton-Cartan geometry {\it prior} to integrating out the massive mode (recall that unlike a Type-II mode, a Type-I mode is canonically conjugate to a massive mode).
We obtain
\begin{multline}
 \mathcal{L} = \sqrt{g} \big( -2 \beta v^\mu \partial_\mu \phi - \tfrac{h^{\mu\nu}}{2m}[\partial_\mu \phi \partial_\nu \phi +\partial_\mu \beta \partial_\nu \beta ]\\ - 2mC^2(x) \beta^2 \big),
\end{multline}
where $c^2 = \rho^{2/d} C^2$ is the speed of sound of the Goldstone mode (the factor of density essentially accounts for the units of $h^{\mu\nu}$).
If we integrate out the massive mode $\beta$ in the limit where we can neglect the dispersion (i.e. at long wavelengths), we recover the Type-I relativistic Lagrangian
\begin{equation}
    \mathcal L_{\mathrm{eff}} = \frac{\sqrt{g}}{2m}\left( \frac{(v^\mu \partial_\mu \phi)^2}{C^2}  - h^{\mu \nu} \partial_\mu \phi \partial_\nu \phi\right).
\end{equation}
From this, we can identify the relativistic metric $\mathcal G_{\mu\nu}$ by observing that this Lagrangian must be of the form in Eq.~\eqref{eq:real-inner-product} such that
\begin{equation}
    \sqrt{-\mathcal G} \mathcal G^{\mu\nu} = \tfrac{\sqrt{g}}{m}\left(\tfrac{v^\mu v^\nu}{C^2} - h^{\mu\nu} \right).
\end{equation}
This yields an equation relating the relativistic metric to the Newton-Cartan object and the gap of the massive mode.
We find
\begin{equation}
    \begin{split}
        \mathcal G_{\mu \nu} & = (m C)^{-\frac2{d-1}} \left( C^2 n_\mu n_\nu - h_{\mu\nu}\right),\\
        \mathcal G^{\mu \nu} & = ( m C)^{\frac2{d-1}} \left(\tfrac{v^\mu v^\nu}{C^2} - h^{\mu\nu}\right),
    \end{split}
\end{equation}
where $d$ is the spatial dimension.
As we can see, the relativistic metric depends crucially on the potential $C(x)$.

This is helpful since, on the one hand, we can easily obtain the stress-energy tensor in the relativistic theory by varying $\delta \mathcal{G}_{\mu\nu}$.
On the other hand, we can use the above formulae to connect this result to the actual stress tensor and energy current/momentum density of the non-relativistic model.
In particular,
\begin{multline}
    \delta \mathcal G^{\mu\nu} =  ( m C)^{\frac2{d-1}} [( v^\mu h^{\nu\lambda} + v^\nu h^{\mu\lambda} - 2\tfrac{v^\mu v^\nu}{C^2} v^\lambda) \delta n_\lambda \\
   + \tfrac{1}{C^2}(v^\mu \delta^\nu_\lambda + v^\nu \delta^\mu_\lambda) \delta u^\lambda + \delta \eta^{\mu\nu}].
\end{multline}
Thus, we can directly relate the relativistic energy-momentum tensor $\mathcal T_{\mu\nu}$ to its non-relativistic counterparts by expanding
\begin{equation}
    \delta S = \int d^{d+1} x \tfrac12 \sqrt{-\mathcal G} \mathcal T_{\mu\nu} \delta \mathcal G^{\mu\nu}
\end{equation}
in terms of the geometric objects in the NC geometry. Doing so, we obtain
\begin{equation}
    \begin{split}
        T^{\mu\nu} & =\tfrac1{m(m C)^{\frac 4{d-1}}} (\delta_\alpha^\mu - n_\alpha v^\mu) \mathcal T^{\alpha\beta}  (\delta_\beta^\nu - n_\beta v^\nu), \\
        \epsilon^\lambda & = \tfrac{1}{m (m C)^{\frac 2{d-1}} }  v^\mu  \tensor{\mathcal T}{_\mu^{\lambda}}, \\
        p_\lambda & = -\tfrac1{mC^2} (\mathcal T_{\lambda \mu} v^\mu - v^\mu \mathcal T_{\mu\nu} v^\nu n_\lambda).
    \end{split}
\end{equation}
where the indices on $\mathcal T^{\mu\nu}$ and $\tensor{\mathcal T}{_\mu^{\lambda}}$ are raised with $\mathcal G^{\mu\nu}$ while all Newton-Cartan objects use the metric $g_{\mu\nu}$.
Ignoring the factors in front of these expressions, one can think of $v^\nu$ as a timelike vector with respect to the metric $\mathcal G_{\mu\nu}$.
In this case, $v^{\mu}$ is directly related to the field of the fluid flow and the object $\mathcal E^\lambda \propto v^\nu \tensor{\mathcal T}{_\mu^{\lambda}}$ is the energy current measured by an observer comoving with that flow (\emph{not} the lab observer).
By the same token $\mathcal P_\lambda \propto \mathcal T_{\lambda\nu} v^\nu$ is the momentum density measured by the comoving observer as well.
Relativistically, these are strictly related $\mathcal E^\lambda = \mathcal G^{\lambda \mu} \mathcal P_\mu$.
However, momentum is imposed by the underlying non-relativistic field theory to be orthogonal to flow $v^\lambda p_\lambda = 0$.
The form of $p_\lambda$ that accomplishes this includes the comoving energy density $v^\mu \mathcal T_{\mu\nu} v^\nu$ and subtracts it off.
Lastly, $T^{\mu\nu}$ is directly related to $\mathcal T^{\alpha \beta}$ projected to live only on spatial slices $n_\mu T^{\mu\nu} = 0$, again as imposed by the underlying non-relativistic theory.

In effective, relativistic, analog systems, there is a preferred (lab) frame that is captured by the Newton-Cartan geometry (in particular $n_\mu$ specifies the lab frame's ``clock'').
This preference is hidden in the high frequency dispersion of the type-I modes and, as we have shown here, results in non-trivial momentum currents and stress-tensors.

As a particular example, a Hawking flux against the flow in an analog system should result in a real energy and momentum current away from the analog black hole.
Far from the horizon (considering the effective 1+1D problem where the other two spatial dimensions are trivial) we obtain
\begin{equation}
    \mathcal T_{\mu\nu} =\begin{pmatrix}
      \mathcal T_{\mathrm{H}} &  -\mathcal T_{\mathrm{H}} \\
       -\mathcal T_{\mathrm{H}} &  \mathcal T_{\mathrm{H}}
    \end{pmatrix},
\end{equation}
for a constant $\mathcal T_{\mathrm{H}}$ \cite{daviesEnergymomentumTensorEvaporating1976} (for the radiation flowing to $+\infty$).
If we apply this to the above, and assume that at $+\infty$ we have no velocity so that $v^\mu = (v^0, \mathbf{0})$ and a flat $h^{ij} =  \delta^{ij}/h_0^{1/3}$, we have
\begin{equation}
\begin{split}
    T^{xx} & = \tfrac1{m h_0^{2/3}} \mathcal T_{\mathrm H}, \\
    \epsilon^\lambda & = \tfrac{v^0}{m}\mathcal T_{\mathrm{H}} \left[ \tfrac{(v^0)^2}{C^2},h_0^{-1/3},0,0\right],\\
    p_\lambda & = \tfrac{v^0}{m C^2} \mathcal T_{\mathrm{H}} [0,1,0,0].
    \end{split}
\end{equation}
Importantly, we see that there is a finite energy current $\epsilon^1$ and momentum $p_1$ away from the horizon; there is no $p_0$ component due to the constraint $p_\mu v^\mu = 0$.
While related to what is computed relativistically, these quantities are not exactly the same.

\section{\label{sec:conclusion}Discussion and Conclusions}

The primary result of this paper is establishing the connection between the different types of Goldstone modes and different types of analog spacetimes, as summarized in Table~\ref{tab:Key_results}.
This is done by revisiting the proof of the non-relativistic Goldstone theorem and allowing for the possibility of an inhomogeneous mean-field solution.
We then find that the conventional Type-I Goldstone modes come equipped with an Einstein-Hilbert metric as appears in general relativity while Type-II Goldstone modes couple to a Newton-Cartan geometry.
The geometry itself is determined by the spacetime dependence of symmetry-breaking mean-field\textemdash inhomogeneous symmetry breaking ultimately produces the non-trivial spacetime metric.
In this work we have restricted ourselves to the case where only the overall $U(1)$ symmetry is inhomogeneously broken.
This corresponds to an overall condensate flow.

Another key result of this paper is establishing the connection between quantum phase transitions and changes in the nature of the spacetime.
To elucidate this, we present a simple model where the analog geometry can be tuned by a single parameter.
This drives a quantum phase transition which accompanies the transition between the Einstein-Lorentz geometry and Newton-Cartan geometry.
As the phase transition is approached, the Hawking radiation produced by an event horizon changes, as encapsulated in Fig.~\ref{fig:totalflux}.
One key result is that the Newton-Cartan geometry exhibits no Hawking radiation, even though all fluid flows are supersonic (the group velocity of Goldstone modes vanishes at long wavelengths).

While Sec.~\ref{sec:model} is a minimal theoretical model, the experimental system that most readily realizes these geometries are spin-1 condensates.
In this case, for the scattering lengths $a_0$ and $a_2$ (for $s$-wave collisions into the spin-0 and spin-2 channels respectively), there are two phases that break the spin SU(2) symmetry: $a_0>a_2$ gives a ferromagnetic phase with one Type-II magnon and $a_0<a_2$ gives a polar phase (antiferromagnetic interactions) with two Type-I magnons.
Upon flow, these two phases naturally realize the two different spacetimes described here.
In fact, ${}^7$Li, ${}^{41}$K, and ${}^{87}$Rb realize the ferromagnetic phase \cite{stamper-kurnSpinorBoseGases2013} with ${}^{87}$Rb specifically already being used for Hawking-like experiments with the phonon mode \cite{steinhauerObservationSelfamplifyingHawking2014,*steinhauerObservationQuantumHawking2016}.
Additionally, ${}^{23}$Na realizes the polar phase and critical spin superflow has been studied \cite{kimCriticalSpinSuperflow2017} (necessary for Hawking-like experiments).
The magnon excitations in these systems can be probed by observing correlations in the spin-density, and the most basic proposal would be to establish the vanishing Hawking radiation in the ferromagnetic phase.
{Though we assume that the spin state is initially homogeneous, while the condensate is flowing, this is a reasonable assumption provided that the condensate can be initially prepared into the homogeneous spin-polarized ground state.
Once this is achieved, accelerating the condensate flow will not produce spin currents and we will obtain the setup we envision in this work~\cite{barnettGeometricalApproachHydrodynamics2009,lamacraftLongwavelengthSpinDynamics2008}. }
The progress in current spinor condensate experiments highlights that these more exotic analog spacetimes may already be in reach.

{On a more abstract level, our work points to the deep connection between the emergent geometry, codified by the objects of the Newton-Cartan geometry, and the superfluid state, characterized by the superfluid density and current.
Indeed, it seems that even in the presence of an enlarged internal symmetry group, such as the $SU(2)\times U(1)$ symmetry of the system we consider here, the spatial variations in the $U(1)$ condensate phase play a special role.
Whereas non-trivial space-time textures of the spin-components can generate extremely interesting non-Abelian synthetic gauge fields (see App.~\ref{app:spin-textures}), only the overall condensate phase can produces a non-trivial analogue spacetime.
In particular, it would be interesting to study how the identification of the Newton-Cartan velocity field $v^\mu$ with the superfluid velocity $\mathbf{v}_s$ possibly leads to novel constraints or techniques for the calculation of transport phenomena in superfluids, essentially expanding upon the framework we have laid out in Sec.~\ref{sec:transport}. }

Finally, by considering the response of the Goldstone modes to variations in the analog geometries, we relate the analog stress-energy-momentum tensor in relativistic geometries directly to their non-relativistic counterpart.
This is summarized by the equations below, which shows how the metric tensor in both analog spacetimes may be constructed from the underlying geometric objects of the Newton-Cartan geometry along with an additional field $C = C(x)$:
\begin{equation}
    \begin{split}
        g_{\mu\nu} & = n_\mu n_\nu + h_{\mu\nu}, \quad \text{Non-relativistic},\\
        \mathcal G_{\mu\nu} & \propto C^2 n_\mu n_\nu - h_{\mu\nu}, \quad \text{Relativistic}.
    \end{split}
\end{equation}
We also provide a direct connection between the energy and momentum currents of an analog relativistic system and the more fundamental Newton-Cartan geometry which describes the lab-frame.

Within spinor Bose-Einstein condensates, there are other phenomena to include such as inhomogeneous broken non-Abelian symmetry (including textures like spiral magnetization, Bloch domain walls, and skyrmions) and synthetic gauge fields.
The construction presented here also considers just the quadratic excitations, but these Goldstone modes realize more complicated nonlinear sigma models for which there is extra \emph{intrinsic} geometry at play and would need to be incorporated into a full theory of these excitations.
This new analog also raises questions of the so-called back-reaction effects of quantum fields on the corresponding analog spacetime.
This has been studied in the relativistic case \cite{fischerDynamicalAspectsAnalogue2007,keserAnalogueStochasticGravity2018}, and the non-relativistic case leaves us with the tantalizing prospect of a system with a dynamical Newtonian gravity.
Finally, while in this work we exclusive focused on the context of flowing spinor Bose-Einstein condensates, the phenomenon should be more general.
An interesting future direction to pursue would be to try and extend these results to include more diverse platforms including electrons in solid-state systems, liquid Helium, superconductors, magnetic systems.
The wide variety of systems which exhibit symmetry-breaking means there is a wide variety of systems which might exhibit this analog spacetime and its consequences.

\begin{acknowledgments}
We would like to thank Gil Refael for crucial discussions which lead to this work.
We also thank Andrey Gromov and Luca Delacr\'etaz for indispensable suggestions.
 This work was supported the U. S. Army  Research Laboratory and the U. S. Army Research Office under
  contract number W911NF1810164, NSF DMR-1613029, and the Simons Foundation (J.B.C. and V.G.).
J.H.W.\ and V.G.\ performed part of this work at the Aspen Center for Physics, which is supported by National Science Foundation grant PHY-1607611. J.B.C.\ and V.G.\ performed part of this work at the Kavli Institute for Theoretical Physics and thank KITP for hospitality and support. J.B.C.\ was supported in part by the Heising-Simons Foundation, the Simons Foundation, and National Science Foundation Grant No. NSF PHY-1748958. J.H.W.\ thanks the Air Force Office for Scientific Research for support.
\end{acknowledgments}

\appendix

\section{Calculating the fluctuation Lagrangian}
\label{app:fluctuations}

In this section, we put all of the algebra and Lagrangian manipulation that we left out of Section~\ref{sec:spacetimes}.

Our starting point is Eq.~\eqref{eq:general-Lagrangian} upon substituting $\Psi = \Psi_0 + \delta\Psi$ where $\Psi_0$ solves the Euler-Lagrange equations Eq.~\eqref{eq:EulerLagrange} and $\delta \Psi$ can be written in terms of broken generators and massive fields Eq.~\eqref{eq:fluctuation-expanded}.

Most of the simplifying algebra comes from $g(\bm \sigma \Psi, \xi) = 0$ and integration-by-parts.
To facilitate the integration by parts, all equalities are be understood to be \emph{up to a full derivative}.
Furthermore, by construction the linear terms cancel, so we keep second-order terms only, indicated by $\stackrel{\mathrm{fluc}}{=}$.

To deal with the term linear in derivatives, we use the object
\begin{equation}
    f \overleftrightarrow{\partial_t} g \equiv f (\partial_t g) - (\partial_t f) g,
\end{equation}
and for simplicity we sometimes replace $\partial_t f$ with $\dot f$ for time derivatives.
We further take advantage of the Einstein summation convention (sum over indices is implied).
The first term we investigate is
\begin{multline}
    \tfrac{i}{2}\Psi^\dagger\overleftrightarrow{\partial_t}\Psi \stackrel{\mathrm{fluc}}{=} -\tfrac{i}{2} \Psi_0^\dagger (\bm\sigma \overleftrightarrow{\partial_t} \bm\sigma) \Psi_0 \\+ i(-\Psi_0^\dagger \bm\sigma \dot \xi + \xi^\dagger \dot{\bm\sigma} \Psi_0 +  \xi^\dagger \bm\sigma \dot\Psi_0) \\- \tfrac{i}2 \Psi_0^\dagger \bm\sigma^2\dot\Psi_0 + \tfrac{i}2 \dot\Psi_0^\dagger \bm\sigma^2\Psi_0.
\end{multline}
Performing integration-by-parts on the $\Psi_0^\dagger \bm\sigma \dot \xi$ term, we get
\begin{multline}
    \tfrac{i}{2}\Psi^\dagger\overleftrightarrow{\partial_t}\Psi \stackrel{\mathrm{fluc}}{=} -\tfrac{i}{2} \Psi_0^\dagger (\bm\sigma \overleftrightarrow{\partial_t} \bm\sigma) \Psi_0 + i(\Psi_0^\dagger \dot{\bm\sigma} \xi + \xi^\dagger \dot{\bm\sigma} \Psi_0) \\
    + i\dot\Psi_0^\dagger (\tfrac12 \bm\sigma^2\Psi_0 + \bm\sigma \xi )
    - i (\tfrac12\Psi_0^\dagger \bm\sigma^2 - \xi^\dagger \bm\sigma)\dot\Psi_0     { + \tfrac{i}2 \xi^\dagger \overleftrightarrow{\partial_t}\xi}. \label{eq:wzmterm}
\end{multline}
\begin{widetext}
The kinetic energy term takes the form
\begin{multline}
  \partial_j \Psi^\dagger \partial_j \Psi \stackrel{\mathrm{fluc}}{=} -\tfrac12 \partial_j \Psi_0^\dagger \bm\sigma^2 \partial_j \Psi_0 - \partial_j \Psi_0^\dagger \bm \sigma \partial_j \bm \sigma \Psi_0 - \partial_j \Psi_0^\dagger \bm \sigma \partial_j \xi  -\tfrac12 \partial_j \Psi_0^\dagger \bm\sigma^2 \partial_j \Psi_0 - \Psi_0^\dagger (\partial_j \bm \sigma) \bm\sigma \partial_j \Psi_0 + \partial_j \xi^\dagger \bm \sigma \partial_j \Psi_0 \\
  - \Psi_0^\dagger \partial_j \bm \sigma \partial_j \bm \sigma \Psi_0 + \partial_j \xi^\dagger \partial_j \xi - \Psi_0^\dagger \partial_j \bm\sigma \partial_j \xi + \partial_j \xi^\dagger \partial_j\bm\sigma \Psi_0.
\end{multline}
We perform integration by parts on the two instances of $-\tfrac12 \partial_j \Psi_0^\dagger\bm\sigma^2 \partial_j\Psi_0$ above in opposite ways to obtain
\begin{multline}
  \partial_j \Psi^\dagger \partial_j \Psi \stackrel{\mathrm{fluc}}{=}
  \tfrac12 \nabla^2 \Psi_0^\dagger \bm\sigma^2 \Psi_0
  - \tfrac12 \partial_j \Psi_0^\dagger \bm \sigma \partial_j \bm \sigma \Psi_0
  + \tfrac12 \partial_j \Psi_0^\dagger (\partial_j\bm\sigma) \bm \sigma \Psi_0
  +\tfrac12 \Psi_0^\dagger \bm\sigma^2 \nabla^2 \Psi_0
  - \tfrac12\Psi_0^\dagger (\partial_j \bm \sigma) \bm\sigma \partial_j \Psi_0
  + \tfrac12\Psi_0^\dagger \bm\sigma (\partial_j \bm\sigma) \partial_j\Psi_0\\
    - \partial_j \Psi_0^\dagger \bm \sigma \partial_j \xi
  + \partial_j \xi^\dagger \bm \sigma \partial_j \Psi_0
  - \Psi_0^\dagger \partial_j \bm \sigma \partial_j \bm \sigma \Psi_0
  + \partial_j \xi^\dagger \partial_j \xi
  - \Psi_0^\dagger \partial_j \bm\sigma \partial_j \xi
  + \partial_j \xi^\dagger \partial_j\bm\sigma \Psi_0.
\end{multline}
If we further use integration by parts on $-\partial_j\Psi_0^\dagger \bm\sigma \partial_j\xi$ and $\partial_j\xi^\dagger \bm \sigma \partial_j \Psi_0$, we obtain (after some reordering)
\begin{multline}
  \partial_j \Psi^\dagger \partial_j \Psi \stackrel{\mathrm{fluc}}{=}
    - \Psi_0^\dagger \partial_j \bm \sigma \partial_j \bm \sigma \Psi_0
   - \tfrac12 \partial_j \Psi_0^\dagger \bm \sigma \partial_j \bm \sigma \Psi_0
  + \tfrac12 \partial_j \Psi_0^\dagger (\partial_j\bm\sigma) \bm \sigma \Psi_0
  - \tfrac12\Psi_0^\dagger (\partial_j \bm \sigma) \bm\sigma \partial_j \Psi_0
  + \tfrac12\Psi_0^\dagger \bm\sigma (\partial_j \bm\sigma) \partial_j\Psi_0\\
    + \partial_j \xi^\dagger \partial_j \xi
  + \partial_j \Psi_0^\dagger \partial_j\bm \sigma \xi
  -  \xi^\dagger \partial_j\bm\sigma \partial_j \Psi_0
  - \Psi_0^\dagger \partial_j \bm\sigma \partial_j \xi
  + \partial_j \xi^\dagger \partial_j\bm\sigma \Psi_0 \\
  + \nabla^2 \Psi_0^\dagger (\tfrac12 \bm\sigma^2 \Psi_0 +\bm\sigma \xi)
   +(\tfrac12 \Psi_0^\dagger \bm\sigma^2 - \bm\sigma\xi) \nabla^2 \Psi_0. \label{eq:kineticterm}
\end{multline}
We observe that, along with Eq.~\eqref{eq:potential-expand}, the equation of motion cancels the last lines in Eqs.~\eqref{eq:wzmterm} and \eqref{eq:kineticterm} with the first line of Eq.~\eqref{eq:potential-expand}.

All together, we can combine these equations to get the full fluctuation Lagrangian
\begin{multline}
  \mathcal L \stackrel{\mathrm{fluc}}{=}
   -\tfrac{i}{2} \Psi_0^\dagger (\bm\sigma \overleftrightarrow{\partial_t} \bm\sigma) \Psi_0 + i(\Psi_0^\dagger \dot{\bm\sigma} \xi + \xi^\dagger \dot{\bm\sigma} \Psi_0)
    { + \tfrac{i}2 \xi^\dagger \overleftrightarrow{\partial_t}\xi} -\tfrac1{2m}[
    - \tfrac12 \partial_j \Psi_0^\dagger \bm \sigma \partial_j \bm \sigma \Psi_0
  + \tfrac12 \partial_j \Psi_0^\dagger (\partial_j\bm\sigma) \bm \sigma \Psi_0
  - \tfrac12\Psi_0^\dagger (\partial_j \bm \sigma) \bm\sigma \partial_j \Psi_0\\
  + \tfrac12\Psi_0^\dagger \bm\sigma (\partial_j \bm\sigma) \partial_j\Psi_0
    + \partial_j \xi^\dagger \partial_j \xi
  + \partial_j \Psi_0^\dagger \partial_j\bm \sigma \xi
  -  \xi^\dagger \partial_j\bm\sigma \partial_j \Psi_0
  - \Psi_0^\dagger \partial_j \bm\sigma \partial_j \xi
  + \partial_j \xi^\dagger \partial_j\bm\sigma \Psi_0
  - \Psi_0^\dagger \partial_j \bm \sigma \partial_j \bm \sigma \Psi_0
   ] \\
 - \frac12 \xi^*_a  \left.\frac{\partial^2 V}{\partial \Psi^\dagger_a \partial \Psi^\dagger_b}\right|_0 \xi^*_b -  \xi^*_a \left.\frac{\partial^2 V}{\partial \Psi^\dagger_a \partial \Psi_b}\right|_0 \xi_b - \frac12 \xi_a \left.\frac{\partial^2 V}{\partial \Psi_a \partial \Psi_b}\right|_0 \xi_b . \label{eq:fluc_lagrang_gen}
\end{multline}
We can now expand our fluctuations in terms of their fields $\bm\sigma \Psi_0 = \theta_n \sigma_n \Psi_0$ and $\xi = \beta_n \xi_n$, and we obtain
\begin{multline}
    \mathcal L \stackrel{\mathrm{fluc}}{=}
    -\tfrac{i}{2}\Psi_0^\dagger[\sigma_m,\sigma_n]\Psi_0 \theta_m\partial_t \theta_n
    +\tfrac1{4m} \theta_m \partial_j \theta_n(\partial_j \Psi_0^\dagger [\sigma_m,\sigma_n]\Psi_0 -  \Psi_0^\dagger [\sigma_m,\sigma_n]\partial_j\Psi_0) \\
    +i \beta_n \partial_t \theta_n (\Psi_0^\dagger \sigma_m \xi_n + \xi_n^\dagger \sigma_m \Psi_0)
    + \tfrac{1}{2m}\beta_m\partial_j\theta_n(\xi_m^\dagger \sigma_n\partial_j\Psi_0 - \partial_j \Psi^\dagger_0 \sigma_n \xi_m + \Psi_0^\dagger\sigma_n \partial_j\xi_m - \partial_j \xi^\dagger_m \sigma_n \Psi_0) \\
    +\tfrac{1}{2m}\Psi_0^\dagger \sigma_n\sigma_m \Psi_0 \partial_j\theta_n\partial_j\theta_m \\
    +\tfrac{i}2 \beta_m \partial_t\beta_n(\xi_m^\dagger \xi_m - \xi_n^\dagger\xi_m)
    +\tfrac{i}2 \beta_n\beta_m (\xi_m^\dagger \partial_t \xi_n - \partial_t\xi_m^\dagger \xi_n)
    +\beta_m\partial_j\beta_n(\xi_n^\dagger \partial_j\xi_m + \partial_j\xi_m^\dagger \xi_n) \\
    -\tfrac1{2m} \xi_n^\dagger \xi_m \partial_j\beta_m\partial_j\beta_n
    -\tfrac12\beta_n\beta_m\left[\xi_{n}^\dagger\left.\frac{\partial^2 V}{\partial\Psi^\dagger\partial\Psi^\dagger}\right|_0 \xi_{m}^* + \xi_n^T \left.\frac{\partial^2 V}{\partial\Psi\partial\Psi}\right|_0 \xi_m + 2 \xi_m^\dagger \left.\frac{\partial^2 V}{\partial\Psi^\dagger\partial\Psi}\right|_0\xi_n\right]. \label{eq:entire-fluctuation-Lagrangian}
\end{multline}
The first three lines of Eq.~\eqref{eq:entire-fluctuation-Lagrangian} lead to the Lagrangian presented in the text Eq.~\eqref{eq:partial-fluctuation-Lagrangian} while the last two lines represent the massive modes neglected in the main text.
\end{widetext}

One can then easily check that once the full Lagrangian in Eq.~\eqref{eq:full-fluc-Lagrangian} is derived that the massive modes conjugate to Goldstone modes no longer have the term that goes as $\beta_m\partial_\mu\beta_n$, only keeping the kinetic term and mass matrix (which we diagonalize to find the type-I basis states).

{
\subsection{Spin textures}
\label{app:spin-textures}

To incorporate spin textures into this theory, we need to make a looser assumption on our mean-field state. 
In this situation, we allow for the broken generators to depend on space and this can be easily accomplished with the introduction of a new field $A_\mu(x)\in \mathfrak g(x)/(\mathfrak u(1)\times \mathfrak h(x))$ such that the mean field satisfies
\begin{multline}
    \Psi_0(x) = \sqrt{\rho(x)} e^{i\vartheta(x)} \chi(x), \quad \chi(x)^\dagger \chi(x) = 1, \\ \partial_\mu \chi(x) = A_\mu(x) \chi(x),
\end{multline}
and in terms of the (spatially-dependent) broken generators $A_\mu(x) = \theta_j(x) \sigma_j(x)$. 
As a concrete example, consider a simple spin-wave in two-dimensions such that $\chi(x,y) = e^{i S_x x} e^{i S_y y} \chi_0$, then $A_x(x,y)= i S_x$ and $A_y(x,y) = \cos(x) i S_y - \sin(x) i S_z$.
$A_\mu(x)$ represents infinitesimal spin-rotations in the $dx^\mu$ direction. 
The removal of the $U(1)$ subgroup from $G/H$ corresponds to the generator $i \mathbb{I}$ which is accounted for with the phase $\vartheta(x)$.
We separate out phase and density since this generator is always broken for a nonzero mean-field, and its explicit relation to fluid flow leads to implications for Galilean boosts; in particular, the gradient of $\vartheta(x)$ is exactly related to fluid velocity.
% In terms of general, linear operators that can be applied to a field, those can be written in terms of GL($N,\mathbb C$) = SL($N,\mathbb C) \rtimes \mathbb C$ and the density and phase fluctuations are contained within the multiplicative group $\mathbb C$ while all spin-texture fluctuations live within SL($N,\mathbb C$), and it is within this group we will start to see things like a covariant derivative appear.

The field $A_\mu(x)$ helps us to determine the spatial dependence of the broken generators.
To understand this, the definition of an unbroken generator is that it must annihilate the mean-field 
\begin{equation}
    \tau_a(x)\chi(x) = 0,
\end{equation}
and by taking a derivative, one can show that $\tau_a(x)$ (and hence $\sigma_b(x)$) obey the unitarity-preserving differential equations 
\begin{equation}
\begin{aligned}
    \partial_\mu \tau_a(x) & = [A_\mu(x), \tau_a(x)], \\
    \partial_\mu \sigma_b(x) & = [A_\mu(x), \sigma_b(x)] .
\end{aligned}
\end{equation}
We further need to define the (spatially-independent) structure constants, for which we have
\begin{equation}
\begin{aligned}[]
    [\tau_a(x),\tau_b(x)] & = h^c_{ab} \tau_c(x), \\
    [\sigma_a(x),\tau_b(x)] & = \tilde g^c_{ab} \tau_c(x) + g^c_{ab} \sigma_c(x),  \\
    [\sigma_a(x),\sigma_b(x)] & = \tilde f^c_{ab} \tau_c(x) + f^c_{ab}\sigma_c(x).
\end{aligned}
\end{equation}

To define a covariant derivative for the Goldstone modes, we write  $A_\mu(x)=A_\mu^n(x)\sigma_n(x)$ and $\bm \sigma = i\theta_0(x) + \theta_n(x) \sigma_n(x)$ (sum over $n$, from 1 to the number of broken generators) and evaluate
\begin{equation}
    (\partial_\mu {\bm \sigma}) \Psi_0 = (i\partial_j\theta_0 + [\nabla_j\theta]_n\sigma_n )\Psi_0
    %(\{ \delta^m_n \partial_\mu + A_\mu^b(x) f^m_{bn}\} \theta_n) \sigma_m \chi.
\end{equation}
where we have defined the covariant derivative
\begin{equation}
     [\nabla_\mu]^m_n \equiv \delta^m_n \partial_\mu + A_\mu^b(x) f^m_{bn},
\end{equation}
and we use the shorthand $[\nabla_\mu \theta]_m \equiv [\nabla_\mu]_m^n \theta_n$.

This gives us enough to deal with the first term in Eq.~\eqref{eq:fluc_lagrang_gen}
\begin{equation}
  -\tfrac{i}{2}\Psi_0^\dagger (\bm\sigma \overleftrightarrow{\partial_t}\bm \sigma) \Psi_0 = -\tfrac{i}{2}\rho \chi^\dagger[\sigma_m,\sigma_n]\chi\,  \theta_m [\nabla_t \theta]_n .
\end{equation}
Just as before, the matrix $\Lambda_{mn}\equiv-\frac i2 \chi^\dagger [\sigma_m,\sigma_n] \chi$ is playing a central role, but the existence of the non-abelian covariant derivative makes it impossible to use it to label the fields $\theta_n$ as type-I or type-II without further structure (we return to this below).
Furthermore, $\Lambda_{mn} = f^0_{mn}$ is spatially independent.

Moving on, we can evaluate the following terms in the Lagrangian Eq.~\eqref{eq:fluc_lagrang_gen}
\begin{multline}
  - \tfrac12 \partial_j \Psi_0^\dagger \bm \sigma \partial_j \bm \sigma \Psi_0
  + \tfrac12 \partial_j \Psi_0^\dagger (\partial_j\bm\sigma) \bm \sigma \Psi_0 + \mathrm{c.c.} =  \\ 2 i \rho \partial_j \vartheta \theta_n[\nabla_j\theta]_m \chi^\dagger [\sigma_n,\sigma_m] \chi - 2 \rho \theta_n [\nabla_j \theta]_m (\nu_b f^b_{nm} A^b_j) \\+ \rho \theta_n\theta_q A^b_j \tilde f^m_{bq} A^c_j g^n_{cm} \nu_n 
\end{multline}
where we have defined $\nu_n>0$ via $\psi_0^\dagger \{\sigma_n, \sigma_m\} \psi_0 = -2 \nu_n \delta_{nm}$ [this is just the inner product we defined in Eq.~\eqref{eq:inner-product}].
While the distinction between mode types is less clear, we can still use $\psi_0^\dagger[\sigma_n,\sigma_m]\psi_0$ to break up \emph{generators} into Type-I and Type-II just as we did in Sec.~\ref{sec:nonrel-goldstone-proof}.
For Type-I generators $\nu_n = \mu_n$ and for Type-II $\nu_n=\lambda_n$ [see Eq.~\eqref{eq:generator_norms}].
Next, we note that
\begin{equation}
    -\Psi_0^\dagger \partial_j \bm \sigma \partial_j \bm \sigma \Psi_0 = \rho \partial_j \theta_0 \partial_j \theta_0 + \rho [\nabla_j \theta]_n [\nabla_j\theta]_n \chi^\dagger \bm \sigma_n \bm\sigma_n \chi.
\end{equation}
Lastly, we need to take into account the terms that couple massive modes with Goldstone modes.
In this situation, it is useful to begin to separate Type-I generators from Type-II 
\begin{widetext}
\begin{multline}
    -\xi^\dagger \partial_j \bm \sigma \partial_j \Psi_0 + \partial_j \Psi_0^\dagger \partial_j \bm \sigma \xi + \partial_j \xi^\dagger \partial_j \bm \sigma \Psi_0 - \Psi_0^\dagger \partial_j \bm \sigma \partial_j \xi
    = -4\rho \partial_j \vartheta \beta_0 \partial_j \theta_0 
    + 2 \rho \beta_n \partial_j \theta_0(\chi^\dagger A_j \sigma_n^\mathrm{I} \chi + \chi^\dagger \sigma_n^\mathrm{I} A_j \chi) 
    \\
    + 2 \rho \beta_0 [\nabla_j \theta]_n (\chi^\dagger A_j \sigma_n \chi + \chi^\dagger \sigma_n A_j \chi) 
    -2 \rho \partial_j \vartheta \beta_n [\nabla_j \theta]_m ( \chi^\dagger \sigma_n^\mathrm{I} \sigma_m \chi  + \chi^\dagger \sigma_m \sigma_n^\mathrm{I} \chi)
    + i \rho [\nabla_j \beta]_n [\nabla_j \theta]_m \chi^\dagger [\sigma_n,\sigma_m]\chi
    \\ 
    - i \rho \beta_n [\nabla_j \theta]_m (\chi^\dagger A_j \sigma_n^\mathrm{I} \sigma_m \chi + \chi^\dagger \sigma_m \sigma_n^\mathrm{I} A_j \chi + \chi^\dagger A_j \sigma_m \sigma_n^\mathrm{I}\chi + \chi^\dagger \sigma_n^\mathrm{I}\sigma_m A_j \chi).
\end{multline}
We can add all of these terms together and use $-\frac{i}2 \chi^\dagger [\sigma_n,\sigma_m]\chi = \lambda_n \delta_{m \bar n}= \Lambda_{nm}$ to obtain our effective Lagrangian for Goldstone modes
\begin{multline}
    \mathcal L_{\mathit{eff}} = \rho \frac{1}2 \theta_n \Lambda_{nm}[D_t \theta]_{m} - \frac{1}{2m} \rho [\nabla_j \theta]_n [\nabla_j \theta]_n - \frac{1}{2m} \rho [\nabla_j \beta]_n [\nabla_j \beta]_n - \rho \mathcal M_{mn}^2 \beta_n \beta_m - \rho \mathcal F_{nm}^2 \theta_n\theta_m + 2\rho \beta_0 D_t \theta_0 \\ - \frac{1}{2m} \rho \nabla_j \theta_0 \nabla_j \theta_0 + 2\rho \mu_n \beta_n [D_t \theta]_n
     -\frac{1}{2m} \rho \theta_n [\nabla_j \theta]_m\Omega_{j,nm}  
   + \frac{2}{m} \rho \beta_0 A_j^n[\nabla_j\theta]_n\nu_n + \tfrac{2}{m} \rho \beta_n A_j^n \partial_j \theta_0 \mu_n \\ -\frac{\rho}{2m} \beta_n [\nabla_j \theta]_m X_{j,nm} - \frac1m\rho f^{q}_{bn} \beta_q A^b_j \Lambda_{nm}[\nabla_j\theta]_{m},
\end{multline}
where $\Omega_{j,nm} = -2 \nu_b f^b_{nm} A_j^b$,  
$$X^j_{nm} = - i(\chi^\dagger A_j \sigma_n^\mathrm{I} \sigma_m \chi + \chi^\dagger \sigma_m \sigma_n^\mathrm{I} A_j \chi + \chi^\dagger A_j \sigma_m \sigma_n^\mathrm{I}\chi + \chi^\dagger \sigma_n^\mathrm{I}\sigma_m A_j \chi), $$
and
$\mathcal F^2_{nm} = A^b_j \tilde f^m_{bq} A_j^c g_{cm}^n \nu_n + A^b_j \tilde f^m_{bn} A^c_j g^q_{cm} \nu_n,$
with no sum over $n$ on the right.
We can now introduce Newton-Cartan geometry as we did before, and by absorbing factors of the density into terms appropriately, the Lagrangian takes the form
\begin{multline}
\mathcal L_\mathit{eff} = n_0 \sqrt{h} (2\beta_0 v^\mu \partial_\mu \theta_0 - \frac{h^{\mu\nu}}{2m} (\partial_\mu \theta_0 \partial_\nu \theta_0+\partial_\mu \beta_0 \partial_\nu \beta_0) \\ +2\rho \mu_n \beta_n v^\mu [\nabla_\mu \theta]_n + \theta_n \Lambda_{nm} v^\mu[\nabla_\mu \theta]_m - \frac{1}{2m} h^{\mu\nu}\{[\nabla_\mu \theta]_n[\nabla_\nu  \theta]_m + [\nabla_\mu \beta]_n[\nabla_\nu  \beta]_m )\}-\tilde{\mathcal M}_{mn}^2 \beta_n \beta_m \\- \tilde{\mathcal F}_{mn}^2 \theta_n \theta_m
-\frac{h^{\mu\nu}}{2m} \theta_n \Omega_{\mu,nm} [\nabla_\nu\theta]_m + \frac{2h^{\mu\nu}}{m} \beta_0 A^n_{\mu} [\nabla_\nu \theta]_n \nu_n - \frac{h^{\mu\nu}}{2m} \beta_n X_{\mu,nm} [\nabla_\nu \theta]_m - \frac{h^{\mu\nu}}{m} f^q_{bn} \beta_q A_\mu^b \Lambda_{nm} [\nabla_\nu\theta]_m).
\end{multline}
There are now new fields that encode the effect of the spin texture. 
The major structural difference though is the introduction of a covariant derivative $\nabla_j$ with a non-Abelian, artificial gauge field.
This is the usual artificial gauge field discussed in the cold atomic context~\cite{linSyntheticMagneticFields2009} and represents the natural generalization of that concept to analog curved spaces: all derivatives become covariant in the natural way.
We speculate that the extra terms (new fields) might be able to be folded back into a new geometry, especially if we place restrictions on the allowable spin-texture, but we leave that exploration to future work.

Importantly, this shows that the Newton-Cartan formalism can accommodate spin-textures with a defined covariant derivative but at the cost of added fields and masses. 
\end{widetext}

 }
\section{Bogoliubov Theory for Hawking Emission}
\label{app:BogoliubovHawking}

As per Eq.~\eqref{eqn:BdG-EOM}, the magnon field (written in terms of the complexified spinor $\Phi_3(x) = (\zeta, \zeta^*)^T$) obeys the BdG equation 
\begin{equation}
\left[ i\tau_3 \hat{D}_t + \frac{1}{2m\rho}\nabla\cdot \rho\nabla - g_3\rho\left(\tau_0 + \tau_1\right) \right] \Phi_3(x) =0, \label{eq:appendix-bdg}
\end{equation}
written in terms of the co-moving frame material derivative $\hat{D}_t = \partial_t + \mathbf{v}_s \cdot \nabla $.

Before proceeding, there are two properties of this equation that prove useful.
First is the charge conjugation symmetry: if $\Upsilon$ solves Eq.~\eqref{eq:appendix-bdg}, then so does
\begin{equation}
    \overline{\Upsilon} \equiv \tau_1 \Upsilon^*.
\end{equation}
In particular, this is important since the Nambu spinor should obey the self-conjugate property that $\Phi_3 = (\zeta ,\zeta^*)^T = \overline{\Phi}_3$.
Thus, it is important that this is respected by the equations of motion, which we see it is.

Furthermore, provided the density $\rho(x)$ is time independent, we can define the conserved pseudo-scalar product on the solution space 
\begin{equation}
    (\Upsilon_1,\Upsilon_2) \equiv \int d^d r \,\rho(\mathbf{r}) \Upsilon_1^\dagger(\mathbf{r}) \tau_3 \Upsilon_2(\mathbf{r}). \label{eq:inner-product}
\end{equation}
This scalar product has a number of useful features including that the charge conjugation operation changes the sign, so that
\begin{equation}
    (\overline{\Upsilon}_1,\overline{\Upsilon}_2) = - (\Upsilon_2,\Upsilon_1).
\end{equation}
We use this pseudo-inner product to define a notion of norm for solutions.
Because of the $\tau_3$, this is not the usual $L_2(\mathbb{R}^d)$ norm, and in fact is not a norm at all since it is not positive semi-definite.
There are non-trivial negative norm states which we loosely refer to as ``hole-like" states, in contrast to the ``particle-like" solutions with positive norm.
As remarked earlier, hole-like solutions can be related to particle-like solutions by charge conjugation since if $\Upsilon$ has negative norm we find
\[
(\Upsilon,\Upsilon ) < 0 \Rightarrow (\overline{\Upsilon},\overline{\Upsilon}) > 0 .
\]

To proceed further, we utilize  the (assumed) time-independence of the kernel to further separate the solution $\Upsilon(x) = \Upsilon(\mathbf{r},t) $ into energy eigenmodes
\begin{equation}
\Upsilon(x) = \int \frac{d\omega}{2\pi} W_\omega(\mathbf{r})e^{-i\omega t},
\end{equation}
where $W_\omega(\mathbf{r}) = [U_\omega(\mathbf{r}),V_\omega(\mathbf{r})]^T$ is a two-component spinor which obeys the eigenvalue problem
\begin{equation}
\label{eqn:BdG-1D}
\left[\omega + i\mathbf{v}_s \cdot \nabla  + \frac{1}{2m\rho}\nabla\cdot\rho\nabla \tau_3 - g_3\rho\left(\tau_3 + i\tau_2\right) \right]W_{\omega}(\mathbf{r})=0.
\end{equation}
We refer to \cite{curtisEvanescentModesSteplike2019,macherBlackholeRadiationBoseEinstein2009} for more details of solving this system.
What is important for our discussion are the details of the dispersion relation, which are used to analyze the asymptotic scattering states at spatial infinity.

We now focus on the case of a one-dimensional homogeneous flow.
In this case both the momentum $k$ and lab-frame frequency $\omega$ are good quantum numbers and obey the standard Bogoliubov dispersion relation (using that $mc^2 = g_3 \rho$) of
\begin{equation}
    \label{eqn:BdG-dispersion}
    \omega = v_sk \pm \sqrt{c^2 k^2 + \left(\frac{k^2}{2m}\right)^2 } \equiv \omega_{\pm}(k),
\end{equation}
where the last equality is used to define the lab frequency $\omega_{\pm}(k)$.
At a particular frequency $\omega>0$, we may determine which scattering states are available by finding the real momenta $k$ which obey $\omega = \omega_{\pm}(k)$.

Considering a step-like variation in the flow, the flow profile is as given in Eq.~\eqref{eqn:step-flow}.
For $x<0$ and $x>0$ the solutions to the BdG equations are still plane-waves which obey the Bogoliubov dispersion relation, albeit with different parameters $\rho$ and $v$.
These two dispersion relations are shown Figs.~\ref{fig:SuperToSubHawking} and \ref{fig:SuperToSuperHawking} for fixed values of the condensate velocities $|v_l| > |v_r|$ and densities $\rho_l,\rho_r$.

Instead of the lab frame, we may measure frequency with respect to the frame co-moving with the fluid flow.
This is implemented by Doppler shifting to the (positive) comoving frequency
\begin{equation}
    \Omega(k) \equiv \sqrt{c^2 k^2 + \tfrac{k^4}{4m^2}},
\end{equation}
so that $\omega_\pm(k) = v k \pm \Omega(k)$ ($vk$ amounts to a Galilean boost).

For $|v|<c$ (right dispersion in Fig.~\ref{fig:SuperToSubHawking}), there are only two real-momenta at any positive frequency, which correspond to a right- and left-moving quasiparticle.
For $|v|>c$ (left dispersion in Fig.~\ref{fig:SuperToSubHawking}) a new scattering channel opens whereby a wavepacket with negative free-fall frequency [$\omega_-(k)$] may have positive lab-frame frequency $\omega$.

We find the eigenfunctions for the step potential by employing matching equations at the step.
These impose the continuity requirements
\begin{equation}
    \label{eqn:matching}
    \begin{aligned}
    &\left[W_\omega(x)\right]_{x=0^-}^{x=0^+} = 0   \\
    &\left[\rho\partial_x W_\omega(x)\right]_{x=0^-}^{x=0^+} = 0.   \\
    \end{aligned}
\end{equation}
Additionally, we choose them to satisfy $(W_\omega,\overline{W}_\omega) = 0$ and can be normalized such that $(W_\omega,W_\omega) > 0$ if $\omega = \omega_+(k)$ (positive comoving frequency) and $(W_\omega,W_\omega) < 0$ if $\omega = \omega_-(k)$ (negative comoving frequency).

Combining all of this, we can express the full solution in terms of positive-frequency components only via
\begin{multline}
\label{eqn:mode-expansion}
\Phi_3(x,t) = \int_0^{\infty} \frac{d\omega}{2\pi} \sum_\alpha \bigg[A(W_{\omega\alpha}) W_{\omega\alpha}(x)e^{-i\omega t}  \\
+ A^*(W_{\omega\alpha}) \overline{W}_{\omega\alpha}e^{+i\omega t}\bigg],
\end{multline}
where the $A(W_{\omega,\alpha})$ are the Fourier coefficients of the expansion and $\alpha$ is a set of quantum numbers which are used to label the different degenerate modes at each energy $\omega > 0$.
At this point, we can second quantize the system and promote $\Upsilon$ to an operator.
In such a case, the operator equation looks like
\begin{multline}
\label{eqn:mode-expansion2}
\hat \Upsilon(x,t) = \int_0^{\infty} \frac{d\omega}{2\pi} \sum_\alpha \bigg[a(W_{\omega\alpha}) W_{\omega\alpha}(x)e^{-i\omega t}  \\
+ a^\dagger(W_{\omega\alpha}) \overline{W}_{\omega\alpha}e^{+i\omega t}\bigg],
\end{multline}
where now $a(W_{\omega\alpha})$ are operators satisfying
\begin{equation}
    [a(W_{\omega\alpha}),a^\dagger(W_{\omega'\alpha'})] = (W_{\omega\alpha},W_{\omega',\alpha'}).
\end{equation}
All $W_{\omega\alpha}$ are orthogonal with respect to this inner product, and so $a(W_{\omega\alpha})$ is either a creation \emph{or} annihilation operator based on the sign of the norm.

The system may be exactly solved when the flow is homogeneous, in which case the momentum $k$ is also a good quantum number.
Assuming a solution of the form
\[
W_{\omega}(x) = w_{k}e^{ikx}
\]
produces the momentum space eigenvalue problem
\begin{equation}
\label{eqn:BdG-momentum}
\left[\omega - vk - \frac{1}{2m}k^2 \tau_3 - g_3\rho\left(\tau_3 + i\tau_2\right) \right]w_{k}=0.
\end{equation}
In principle, the momentum $k$ depends in the energy $\omega$, but we usually suppress this dependence for brevity.

To evaluate $(W_{\omega\alpha},W_{\omega'\alpha'})$, we establish a couple of facts. 
If we let $w_k = [u_k, v_k]^T$, then we have
\begin{equation}
    mc^2 v_k = \left( \pm \Omega(k) - \frac{k^2}{2m} - m c^2 \right) u_k,
\end{equation}
and hence
\begin{equation}
 m^2 c^4 |v_k|^2 = \left\{m^2 c^4 \mp 2\Omega(k)\left[mc^2 + \frac{k^2}{2m} \mp \Omega(k) \right]  \right\}|u_k|^2,
\end{equation}
this relation between $|u_k|^2$ and $|v_k|^2$ allows us to evaluate
\begin{widetext}
\begin{equation}
\begin{split}
    (W_{\omega\alpha},W_{\omega'\alpha'}) & = \pm \Omega(k) \frac{2 \rho}{m^2 c^4} \left[ mc^2 + \frac{k^2}{2m} \mp \Omega(k) \right] |u_k|^2 \delta_{\alpha\alpha'}\delta[k_\alpha(\omega) - k_{\alpha'}(\omega')] \\
    & =  \pm \Omega(k) \frac{2 \rho |v_g|}{m^2 c^4} \left[ mc^2 + \frac{k^2}{2m} \mp \Omega(k) \right] |u_k|^2 \delta_{\alpha\alpha'}\delta(\omega - \omega').
\end{split}
\end{equation}
\end{widetext}
The term in brackets $m c^2 + \frac{k^2}{2m} - \Omega(k) > 0$, so the \emph{sign} of the normalization depends exclusively on whether we have positive ($+\Omega(k)$) or negative ($-\Omega(k)$) comoving frequency.
The terms with negative comoving frequency (or negative norm) are represented by the blue curves in Figs.~\ref{fig:SuperToSubHawking}~and~\ref{fig:SuperToSuperHawking}.

We can now perform the Hawking calculation to determine the Bogoliubov transformation giving rise to excitation production.
This is presented first in Fig.~\ref{fig:SuperToSubHawking}, where we consider a wavepacket moving away from the horizon to $+\infty$ and frequency $\omega$, this is the Hawking mode.
If we trace it back in time, it was related to a scattering process at the horizon itself, so in terms of three other positive frequency modes
\begin{equation}
    W_{\mathrm{H}} = \alpha_R W_{R,1} + \alpha_L W_{L,2} + \beta_L \overline{W}_{L,1},
\end{equation}
where $ W_{\mathrm{H}} $ includes the far propagating right-moving mode along with the evanescent near horizon solution, $W_{R,1}$ is the left-moving mode on the right, and $W_{L,(1,2)}$ are the right-moving modes on the left (counted left-to-right in Fig.~\ref{fig:SuperToSubHawking}).
This immediately gives us how to relate the creation operators of the out-vacuum to the in-vacuum
\begin{equation}
    a(W_\mathrm{H}) = \alpha_R a(W_{R,1}) + \alpha_L a(W_{L,2}) + \beta_L a^\dagger(W_{L,1}).
\end{equation}
This implies that for $W_H$ at a particular frequency $\omega$, we can find the number of Hawking modes leaving the horizon by considering the expectation value
\begin{equation}
    \braket{0_{\mathrm{in}}| a(W_\mathrm{H})^\dagger a(W_\mathrm{H}) | 0_\mathrm{in}} = |\beta_L|^2 (W_{L,1}, W_{L,1}).
\end{equation}
With the proper normalization and putting back in the dependence on frequency, the number of particles leaving the horizon at frequency $\omega$ is
\begin{equation}
  N(\omega) = |\beta_L(\omega)|^2 \frac{(W_{L,1}(\omega),W_{L,1}(\omega))}{(W_{H}(\omega),W_{H}(\omega))}. \label{eq:Nomega1}
\end{equation}

This same analysis can be done for the supersonic-to-supersonic case presented in Fig.~\ref{fig:SuperToSuperHawking}.
For lack of a better term, we call the region where there are multiple positive and negative norm channels the ``super-Hawking'' region.
In this case, we have two modes in the Hawking process that need to be backwards scattered: one positive norm and the other negative norm.
The result of the scattering process is
\begin{equation}
  \begin{split}
    W_{\mathrm H} & = \beta_R \overline W_{R,1}  +  \alpha_R W_{R, 2} + \beta_L \overline W_{L,1} + \alpha_L W_{L,2}, \\
    \overline{W}_{\mathrm H'} & = \alpha_R' \overline W_{R,1}  + \beta_R' W_{R, 2} + \alpha_L' \overline{W}_{L,1} + \beta_L' W_{L,2}.
  \end{split}
\end{equation}
These equations can be similarly related to a Bogoliubov transformation, and we can find the number of Hawking particles leaving the horizon at frequency $\omega$ by considering
\begin{widetext}
\begin{equation}
  N(\omega) =  |\beta_L(\omega)|^2 \tfrac{(W_{L,1}(\omega),W_{L,1}(\omega))}{(W_{H}(\omega),W_{H}(\omega))} + |\beta_R(\omega)|^2 \tfrac{(W_{R,1}(\omega),W_{R,1}(\omega))}{(W_{H}(\omega),W_{H}(\omega))}
  + |\beta_L'(\omega)|^2 \tfrac{(W_{L,2}(\omega),W_{L,2}(\omega))}{(W_{H'}(\omega),W_{H'}(\omega))} + |\beta_R'(\omega)|^2 \tfrac{(W_{R,2}(\omega),W_{R,2}(\omega))}{(W_{H'}(\omega),W_{H'}(\omega))}. \label{eq:Nomega2}
\end{equation}
Despite there being more terms, there is generally less of a Hawking flux due to a decoupling of the negative and positive norm channels as we can see in Fig.~\ref{fig:totalflux}.
\end{widetext}

\bibliography{spinorBEC-curvedspace.bib,arxivpapers.bib}

\end{document}